\documentstyle[12pt,aasms4,epsfig]{article}
\long\def\***#1{{\scshape ***#1***}}



\lefthead{PASCHOS}
\righthead{IGM TRANSMISSION}

\begin{document}

\title{A Statistical Analysis of Intergalactic Medium Transmission \\
          Approaching Reionization}
\author{Pascal Paschos \& Michael L. Norman}
\affil{Laboratory for Computational Astrophysics \\ 
   Center for Astrophysics and Space Science, UC San Diego,
  La Jolla, CA 92093\\}

\slugcomment{Astrophysical Journal, submitted}

\begin{abstract} 

We use hydrodynamic cosmological simulations to explore the evolution of
the intergalactic medium (IGM) transmissivity from z=2 through the epoch
of reionization. We simulate a concordance $\Lambda$CDM model in a 9.6 Mpc
box with a comoving spatial resolution of 37.5 kpc. Reionization is
treated in the optically thin approximation using an ultraviolet
background (UVB) that includes evolving stellar and QSO source
populations. In this approximation, ionization bubble overlap is treated
by ramping up the UVB over a finite redshift interval $\Delta z \approx
0.5$, consistent with the inhomogeneous reionization simulations of
Razoumov et al. (2002) for several reionization redshifts. We construct
noiseless synthetic HI Ly$\alpha$ absorption spectra by casting lines of
sight (LOS) through our continuously evolving box and analyze their
properties using standard techniques. Different spectral resolutions are
also studied by convolving full resolution data down to R=36,000 and
R=5,300 depending on redshift. Parametric fits to the data are provided
based on either analytic approximations or straightforward regressions.

We find a smooth evolution of the effective optical depth under a power
law with a slope of $4.16 \pm 0.02$ up to the epoch of reionization. The
smooth profile can also be fitted to the Songaila and Cowie (2002)
parametrization F(g,z). The normalized ionization rate g is then recovered
through our spectra which agrees, within the error of the fit, with the
input ionization rate we used in the simulation. As we cross into the
epoch of reionization, the mean transmitted flux (MTF) and variance to the
mean transmitted flux sharply deviate from a smooth evolution. However,
the simultaneous sharp increase of the variance and sharp decrease of the
mean transmitted flux introduces large margins of error which place a high
degree of uncertainty to the computed optical depth evolution profile. A
distinction between high and low transmission lines of sight shows that
the two subsamples skew the results towards two different directions that
may infer a continuation of a smooth profile or underestimate the global
transmissivity properties during reionization. However, despite the
statistical uncertainty in inferring the reionization profile from
spectra, the end of an opacity phase transition of the IGM correlates well
with the redshift when both the mean and variance of the transmitted flux
rapidly deviate from a smooth profile. Furthermore, we quantify the
relation between the line of sight and the cosmic flux variance, which is
computed from the statistical average of the flux variance along all lines
of sight and conclude that because the latter is a lower bound estimate
due to limitations imposed by our box size, it is a more sensitive tool
than the MTF in mapping reionization. It nonetheless causes the
distribution of mean fluxes along lines of sight to have an increasing
flatness as the redshift increases. We estimate that in our cosmic
realization of reionization and regardless of spectral resolution, an
unobtainable number of lines of sight is needed to yield a normal
distribution of LOS-mean fluxes that would allow an estimate of the MTF
with less than 10\% relative margin of error.

In addition to optical transmission, we compare the predicted dark gap
length distribution with observations. We show that this statistic is
sensitive to spectral resolution at reionization redshifts, but overall in
agreement with results by Songaila and Cowie (2002). Finally, we derive a
positive correlation between the mean optical depth within a gap and the
size of the gap, which relates "transmission statistics" to "dark gap
statistics" in high redshift studies of the IGM.

\end{abstract}


\section{Introduction} \label{intro}

The detection of quasars at $z \geq 6$ (Becker et al. 2002; Fan et al. 2002, 2003; Hu et al. 2002)
suggests that the intergalactic medium is highly ionized by $z=6$ and therefore reionization began at
redshifts $z>6$ due to UV emitting source other than quasars (QSOs). Quasars are not considered to be such
sources because their comoving number density decreases rapidly at $z>4$ (Shapiro \& Giroux 1987; Madau,
Haardt \& Rees 1999).  By virtue of the WMAP results (Bennett et al. 2003) an era of reionization is
associated with the epoch of early star formation at $z \approx 17$. The number of massive PopIII stars
forming then produces sufficient UV photons to at least partially ionize the IGM (Barkana \& Loeb 2001,
2003;  Haiman \& Holder 2003). In addition, the discovery of a large number of Lyman Break Galaxies (LBG)
(Steidel et al. 2003) at redshifts ($z>3$) could also explain the completion of reionization by $z \approx
6$ due to the population of proto-galaxies that form earlier ($z<9$). In that case, massive galactic type
stars, which are abundant at $z>6$ (Madau et al. 1999), would be the ionizing sources of a
galaxy-dominated UV background (Haenhelt et al. 2001). Simulations can be "fine-tuned" to describe one or
the other scenario (Razoumov et al. 2001; Giardi et al. 2003; Sokasian et al. 2003). Regardless of whether
such results will become the standard accepted theory in the study of reionization one conclusion has
gained substantial confidence. Hydrogen reionization was most likely caused by a soft, stellar type UV
radiation field with a large softness index ($S = \frac{\Gamma_{HI}}{\Gamma_{HeII}} > 100$) between
redshifts $z=6-15$.

The large optical depth due to electron scattering detected by the WMAP places the beginning of
reionization much earlier than the highest redshift Ly$\alpha$ emitter (SDSS 1148+5251) detected to date
in ground based observations at $z = 6.56$ (Hu et al. 2002). If the Universe was permanently globally
reionized after $z=15$ then the Ly$\alpha$ opacity evolution is expected to be smooth down to small
redshifts (Songaila 2004). However, Lyman-$\alpha$ forest observations in the spectrum of SDSS~1030+0524
(z=6.28) shows an optical depth trough at $z=6.05$ (Becker et al. 2002, Songaila \& Cowie 2002; Fan et al.
2002). The lack of transmitted flux has been interpreted as the detection of the reionization tail.
However, Songaila (2004) has shown that the previous conclusion was based on an incorrect conversion of
Ly$\beta$ opacities to Ly$\alpha$ opacities. The correct conversions show a smooth evolution of the
optical depth up to $z=6.3$. This suggests that the trough in the SDSS~1030+0524 spectrum might be a 'dark
gap' occurrence due to the large variance in absorption at high redshifts, although it is not clear
whether the line of sight variance (Fan et al. 2002) or the cosmic variance (Songaila 2004) is
responsible.

In this paper we investigate the evolution the mean transmitted flux and flux variance using hydrodynamic
cosmological simulations of the Ly$\alpha$ forest in a concordance $\Lambda$CDM universe. We adopt a
picture of hydrogen reionization that begins at $z \approx 7$ due to the galactic radiative feedback and
is rapidly completed by $z \approx 6.5$. That does not exclude the possibility of recurring reionization
events prior to $z \approx 7$ but does exclude a single reionization instance at $z \approx 15$. In our
picture, the transmitted Ly$\alpha$ flux evolves smoothly from small redshifts up to $z~\approx~6.5$ which
is consistent with the Songaila (2004) result. Knowing the exact moment and shape of reionization in our
simulation allows the investigation of the statistical properties of the Ly$\alpha$ transmission using
synthetic spectra even within the reionization tail. We find that, although the mean transmitted flux
deviates from the extrapolated smooth evolution as it enters the reionization tail, the large scatter in
the transmitted flux from one line of sight to another makes it a poor indicator of reionization.  An
observation could miss the reionization tail or underestimate the redshift at which the tail begins.
Instead we propose that the line of sight variance be used for tracing the reionization tail because it
shows a larger sensitivity to the hydrogen opacity.
 
\section{Simulations} \label{sims}

We have performed a $\Lambda$CDM hydro-cosmological simulation using our Eulerian code Enzo (Bryan \&
Norman 1997; Norman \& Bryan 1998; Bryan et al. 1999) with $\Lambda=0.73$, $h=0.71$ ($H_{o}=100h$
kms$^{-1}$), $\Omega_{m}=0.27$ and $\Omega_{b}=0.04$. The primordial distribution of gas and dark matter
was initialized in a comoving box of $6.816h^{-1}$ Mpc using the linear power spectrum from Eisenstein \&
Hu (1999) with $\sigma_{8}=0.94$. The cosmic, hydrodynamic and ionization evolution of the IGM was then
computed from z=99 to z=2.0. Enzo solves the coupled system of the multispecies, fluid, self-gravity and
ionization equations on an Eulerian comoving grid for the redshift evolution of the 3D distribution of the
gas variables (e.g. temperature, density etc.). The spatial distribution of $256^{3}$ collisionless cold
dark matter particles is also computed at every time-step and is used to calculate the large-scale
gravitational field in the box. The grid resolution of $256^{3}$ cells gives our simulation box a comoving
spatial resolution of $26.62h^{-1}$ kpc.  Bryan et al. (1999) concluded that in order to resolve the
Ly$\alpha$ forest in a numerical simulation one needs at least 40 kpc spatial resolution. For our choice
of the Hubble constant and box size, we resolve comoving scales of ~37 kpc per grid cell.

Enzo uses a volume averaged UV background that evolves with redshift and photo-ionizes the primary cosmic
chemical elements (H \& He). Until recently, Enzo used the mean intensity spectrum computed by Haardt \&
Madau (1996) (HM96) which was based on quasar counts alone. Such a spectrum is insufficient to describe
the ionization and thermal state of the IGM at redshifts $z > 5$. Our current understanding of hydrogen
re-ionization places the onset of re-ionization at redshifts $z > 6.5$ where quasar counts alone cannot
provide the needed flux for the IGM to ionize. One possible source for the missing flux is the population
of dwarf-galaxies, which form in shallow gravity wells between $z=12-7$ and contribute to a soft, stellar
UV background. The other possible source for the ionizing flux due to PopIII objects at redshifts $z > 15$
is not addressed in this paper.

Haardt \& Madau (2001) (HM01) have computed the redshift evolution of the volume averaged UV intensity
which takes into account the evolving populations of both galaxies and QSOs. We have incorporated into
Enzo this radiation background in the form of the frequency-integrated photo-ionization rates, $\Gamma_{i}
= \int_{\nu_{i}}^{\infty} \frac{4 \pi J(\nu,z)}{h \nu} \sigma_{i}(\nu) d \nu $, and photo-heating rates ,
$G_{i} = \int_{\nu_{i}}^{\infty} \frac{4 \pi J(\nu,z)}{h \nu} \sigma_{i}(\nu) (h \nu - h \nu_{i}) d \nu$.
In our notation, i indicates the chemical species (i=HI,HeI,HeII), $\nu_{i}$ the species' ionization
frequency, $\sigma_{i}(\nu)$ the ionization cross section and $J(\nu,z)$ the mean intensity frequency
distribution at redshift z. The photo-ionization and photo-heating rates are used to update the local
chemical abundances and local gas energy respectively.  In Figures~(1) and ~(2) we show parametric fits to
the redshift evolution of the species' photo-ionization and photo-heating rates.  At redshifts $z > 4$ we
note the effect that the galactic contribution has on the UV flux. The QSO contribution diminishes rapidly
at $z \geq 3$ due to the decrease of their comoving number density. However, the galactic component, which
peaks around $z \approx 4$, provides enough ionizing flux to at least ionize HI and HeI (the softness of
the stellar radiation makes it a poor ionizer of HeII which requires a harder spectrum).

A self-consistent calculation will compute the 3D propagation and percolation of the ionization fronts
(I-fronts) emanating from the UV production regions (Shapiro \& Giroux 1987; Abel \& Haehnelt 1999; Gnedin
2000; Razoumov et al. 2002; Giardi et al. 2003;  Sokasian et al. 2003). In that scheme, reionization
begins when the individual I-fronts begin to merge and is completed when the volume filling factor of
ionized matter is $\approx 1$. Prior to the merging, regions of fully neutral hydrogen in the space
between the I-fronts have large column densities and therefore large optical depths. Quantitatively, a
comoving stencil ($\Delta x$)  of neutral hydrogen at the cosmic mean gas density will have a column
density of $N_{HI} \approx 2.6~10^{16}~\Omega_{b}h^{2}(1+z)^{2} \left ( \frac{\Delta x}{1kpc} \right ) $
cm$^{-2}$. At z=7 and for $\Delta x \approx 40$ kpc, the spatial scale that resolves the Ly$\alpha$ forest
(Bryan et al. 1999), we get $N_{HI} \approx 1.5~10^{18}$ cm$^{-2}$ which corresponds to $\tau_{LyC}
\approx 9.3$.  Despite this seemingly large continuum optical depth the I-fronts will eventually burn
through as the number density of the ionizing sources increases and the volume is rarefied by cosmic
expansion and structure formation. Since most of the volume is at overdensities below the cosmic-mean the
I-fronts are rapidly propagating in underdense regions while slowly ionizing denser clumps of neutral
material. This mechanism of reionization accelerates as the I-fronts merge because at that point two or
more UV sources can ionize the same region of space. This accelerated pace is illustrated in the steep
decrease within a very short time ($\Delta z_{reion} = 0.2-0.3$ or 25-40 Myrs) of the Gunn-Peterson
optical depth computed by Razoumov et al. (2002).

The study of reionization using a rising uniform UV background, which is applied at every point in the
cosmic volume, lacks the mechanism of percolating I-fronts and therefore \textit{cannot} self-consistently
simulate the effect of reionization via the merging of ionized regions. In order to proceed we need to
closely emulate the current understanding of \textit{reionization mechanics} in our numerical setup.
First, we need to choose the redshift at which reionization begins which corresponds to the phase of
initial merging of the I-fronts. At that point, we initialize our background UV field to a tiny value. This
approach is not far from reality because, despite the presence of bubbles filled with UV-radiation in the
volume, most of the hydrogen mass is not yet ionized. The effect of our initialization is to begin to
ionize the most underdense regions with a value for the photo-ionization rate that does not yet have an
effect on higher density regions. The last claim is based on the fact that higher gas densities will have
smaller recombination time-scales than lower gas densities and therefore achieve ionization equilibrium
faster. The underdense regions on the other hand may never achieve ionization equilibrium and therefore
their state of ionization is effectively determined by the ionization time-scale ($\Gamma^{-1} \approx
0.16$ Myrs between z=7-6). Ignoring the effects of recombination for small overdensities allows for a
simplified estimate of the time required for a fixed volume element $\delta V$ to be fully ionized. This
is determined by the balance between the number of ionizing photons per unit time and the available
targets per unit volume to be ionized $\delta \tau_{ion} = \frac{n_{HI}\delta V}{\Gamma}$. The last
expression suggests that for a uniform ionization rate the most underdense regions photo-ionize first.  
This 'bottom-up' ionization mechanism emulates the preferred I-front propagation channel which is the
expansion into the underdense IGM.

Our second 'free' parameter is the choice for the time-scale at which we would need to ramp-up the
ionization (and heating) rates to a volume averaged values that is physically valid in the cosmic medium
at high redshifts. In reality, the ramping of the volume-averaged rates is expected due to the rapid
increase of the UV-radiation volume-filling factor as the I-fronts merge. In addition, as the cosmic
volume becomes transparent to more sources of radiation photo-ionization of the IGM accelerates.  Since
the current highest-redshift Ly$\alpha$ emitter lies at z=6.56 (Hu et al. 2002) our choice for the onset
of reionization needs to be $z_{reion} > 6.6$. We adopt the conclusions from the galactic ionization model
by Razoumov et al. (2002) and set $z_{reion}=7$. In addition, we ramp the ionization rates to the HM01
values by $z=6.5$ using an analytic ramp function with a skin-width of $\Delta z_{reion} = 0.3$.

Choosing an earlier initialization redshift for the ionizing background does not have any measurable
effect on the opacity of the IGM at $z < 6.5 $. We demonstrate this in Fig.~3 where we compute the
redshift dependence of the Gunn-Peterson optical depth (Peebles 1965)  $\tau_{GP} \approx
4.4~10^5~\chi_{HI}~(\Omega_{b}h^{2})~h^{-1}~(1+z)^{1.5}$ for five cases using a lower grid resolution
simulation ($128^{3}$ grid cells) of the same box size and substituting for the $\chi_{HI}$ the mean HI
neutral fraction in the volume. Our first case (A0) uses the QSO-only HM96 spectrum where we extrapolate
the fit to the rates beyond z=5.0 in order to initialize the radiation field at $z_{on}=7$. The cases A
through D correspond to the HM01 spectrum and use the ionization rates from Figure~(1). They differ only
in the choice for the redshift of the reionization onset. Our redshift of interest lies at $z \leq 6.5$
where there are current observations. Henceforth, we find no difference in the optical depth evolution
between the four models in that redshift range.  Therefore, we choose $z_{on}=7$ for consistency with
other theoretical and numerical models.

Due to our limited box size, we terminate our calculation at z=2. However, since our focus is the high
redshift Ly$\alpha$ forest, that our computational box contains enough mass power at these redshifts to
make it a representative cosmic realization. A recently completed simulation of a ~54$h^{-1}$ Mpc
cosmology box with a grid resolution of $1024^{3}$ grid cells and $1024^{3}$ dark matter particles will
address the lack of matter power at large scales. We plan to repeat this work for that simulation in order
to determine if the deficiencies of the current simulation has an effect on our conclusions.

\section{Synthetic Spectra} \label{spec}

The method for generating synthetic spectra of the Ly$\alpha$ forest is described in Zhang et al. 1997,
Bryan et al. 1999 and Machacek et al. 2000.  We begin the spectrum calculation by selecting a point in the
volume from which we cast random lines of sight. Along a line of sight (LOS) we integrate the optical
depth of the redshifted Ly$\alpha$ photons that are scattered at the rest frame of reference of an
absorbing grid cell. The optical depth is given by Equation~(1) (Zhang et al. 1997), where $\alpha$ is the
cosmic scale factor and the integration limits are between the redshift of the initial point (equal to the
redshift of the simulation dump) and the redshift at which we wish to measure the Ly-$\alpha$ absorption.

\begin{eqnarray}
\label{optequa}
\tau_{\nu}(z)=\frac{c^2 \sigma_{o}^{i}}{\sqrt{\pi}
\nu_{o}^{i}} \cdot \int_{z}^{z_{o}^{i}} \frac{n^{i}(z')}{b^{i}}
\frac{\alpha^{2}}{\dot{\alpha}} 
\: \times \: exp \{ - [ (1+z')\frac{\nu}{\nu_{o}^{i}}-1+\frac{\upsilon}{c} ]^{2}(\frac{c}{b^{i}})^{2} \} dz'
\end{eqnarray}

In our notation $\sigma_{o}^{i}$ and $\nu_{o}^{i}$ are the resonant cross section and rest-frame
scattering frequency for ion i (e.g. HI or HeII). $b^{i}=\sqrt{\frac{2kT}{m_{i}}}$ is the local thermal
speed and $n^{i}$ is the local proper number ion density. The projected velocity along the LOS at the
local cell, $\upsilon$, is the sum of the projected peculiar velocity plus the Hubble expansion speed. The
input fields used to generate the synthetic spectra are the grid distributions of gas temperature, the
three components of the gas peculiar velocities and the ion proper density. Our redshift integration range
is $\Delta z = 0.1$ which corresponds to the frequency of the simulation dumps. The optical depth
integration assumes that the input fields do not have any comoving evolution between $z_{cube}$ and
$z_{cube}-\Delta z$.

Each line of sight is continuous through redshift space. At the end of integration we store the position
of the last absorbing cell and the direction of the ray and use them at the beginning of the integration
step through the next simulation dump. The size of our simulation box is smaller than the distance
traveled by the Ly$\alpha$ photons in $\Delta z = 0.1$ for all redshifts $z \geq 2$. Therefore each LOS
exits and re-enters the periodic computation box several times before the integration step is completed.
The number of exits and re-entries depends on the redshift due to the volume's proper size increase with
cosmic time. For more details on the subject we refer to Zhang et al. (1997).

The transmitted flux at every redshift point is simply $F_{\nu}=exp(-\tau_{\nu})$.  We call the arithmetic
mean of the transmitted flux along a LOS in a redshift bin [z,z-$\delta z$] the mean value
$\overline{F}(z_{mean})=<F_{LOS}>_{\delta z}$ at $z_{mean} = z-0.5 ~\delta z$. The LOS effective optical
depth at $z=z_{mean}$ is then defined by $\tau_{eff} (z) =-ln(\overline{F}(z))$.  Following Gaztanaga \&
Croft (1999), we define the LOS flux variance, the variance along a line of sight, through
$\sigma_{LOS}^{2} = Var^{2}(\overline{F})=\frac{<F_{LOS}^{2}>} {<F_{LOS}>^{2}}-1$. The normalization by
the mean flux is in analogy to normalizing the density fluctuations to the mean density.  The mean
LOS-variance (MLV) is then equal to $\sigma_{MLV}^{2}=\frac{1}{NLOS}~\sum_{LOS=1}^{NLOS}
\sigma_{LOS}^{2}$.

The mean values from all lines of sight constitute a population of random samples distributed about the
expected value of the mean transmitted flux at that redshift. Therefore, we call the mean transmitted flux
(MTF) in the redshift bin [z,z-$\delta z$] the LOS averaged mean
$MTF(z)=\frac{1}{NLOS}~\sum_{LOS=1}^{NLOS}(<F_{LOS}>_{\delta z})$. The unbiased variance of the MTF at
redshift $z_{mean}$ is computed by $\sigma_{MTF}^{2}=Var^{2}(MTF) = \frac{1}{NLOS-1} \sum_{LOS=1}^{NLOS}
(\frac{<F_{LOS}>}{MTF} - 1)^{2}$, where we follow the same definition convention as with a single line of
sight. Finally we define the total MTF-variance (TMV) as the product $TMV = NLOS \times \sigma_{MTF}^{2}$
(see Section 4.2).

\section{Results} \label{res}

A total of 75 random lines of sight were computed from z=6.6 to z=2.  Spectra synthesized with Voigt
profiles beyond $z>6.6$ did not show any transmission above our flux cutoff ($=e^{-20}=2.1~10^{-9}$).  
Each integration redshift bin of $\Delta z =0.1$ was resolved by 30000 points which results in a maximum
redshift resolution of $R_{z} = 3 \times 10^{5}$. Our spectral resolution then becomes a function of
redshift, $R_{\lambda}=R_{z} \times (1+z_{mean})$ where $z_{mean}$ is the mean of the redshift interval.
Three resolutions are then considered in the present work. The full resolution case (FRES) uses our raw
synthetic spectra. A high resolution (HRES) case convolves our synthetic spectra, in a redshift interval
$\Delta z$ and a mean redshift $z_{mean}$, with a gaussian at the HIRES spectrograph resolution of
$R=36,000$. It is used to compare with low redshift observed data. A low resolution (LRES) case convolves
our synthetic spectra to the ESI (Sheinis et al. \ 2000) resolution of $5,300$. We use the term 'low
resolution' in comparison to the FRES case. Values of $R > 5,000$ are actually quite high resolution for
observed data at $z > 4$. Unless we state otherwise we apply the LRES case only at $z > 4.5$.

In Figure~(4) we show four samples of a synthetic Ly$\alpha$ forest absorption spectrum along a single
random line of sight at $z_{mean} = [3.05,4.05,5.05,6.05]$ in redshift intervals of $\Delta z = 0.1$. At
$z_{mean}=3.05$ and $z_{mean}=4.05$ we plot the transmitted flux in the HRES case. The next two bins
($z_{mean} = 5.05$ \& $z_{mean}=6.05$) are plotted in the LRES case. Between $z \sim 4$ and $z \sim 5$ an
increasing number of low transmission regions (dark gaps) appear in the Ly$\alpha$ forest that, as we
shall show, are underionized high opacity overdensities. The size of the dark gaps increases between $z
\sim 5$ and $z \sim 6$ as the smaller gaps at lower redshifts 'merge' and the amplitude of the high
transmission region decreases.

\subsection{Mean Transmitted Flux Evolution}\label{mtf}

In Figure~(5) we plot the mean transmitted flux (MTF) versus redshift in 30 redshift bins with size
$\delta_{bin} = 0.15$ (solid line) which corresponds to $\delta \lambda = 186$ \AA. Overplotted (diamonds)
are the combined samples of HIRES and ESI data from Songaila (2004) and the individual transmitted fluxes
from each simulated line of sight.  In Figure~(6) the transmitted flux is converted to optical depth. The
red crosses show the evolution of the effective optical depth, $\tau_{eff}=-ln(MTF(z))$. The smoothness of
$\tau_{eff}$ persists in our calculation up to $z_{mean}=6.36$ which is consistent with the conclusion by
Songaila (2004). The solid red line through the red crosses is a power law least squares fit between
$z_{mean}=6.36-2.07$. The last red cross in Figure~(6) lies well within the reionization tail and was not
included in the fit:
\begin{equation}
\tau_{eff}~=~ \,
 2.1^{+0.12}_{-0.11}~\left ( {{1+z} \over {6}} \right )^{4.16 \pm 0.02}
\end{equation}\smallskip\noindent 
The cosmic parameters in our simulation were chosen in order 
to match low redshift observations and indeed the match with observed data at
$z \leq 3.5$ is good. The blue lines in Figure~(7) are a measure of the scatter of the LOS mean
transmitted flux. They correspond to the minimum and maximum LOS mean values
(converted to optical depths) in each redshift interval. 
Within the LOS scatter range we are in decent agreement with the observed values 
across the entire redshift space.

The solid lines in Figure~(6) correspond to the full resolution of our synthetic spectra (FRES). The
spectral resolution does not have an effect on the mean flux value in a redshift interval but it does
alter the scatter of the data about the mean. Low resolution spectra will have smaller scatter about the
mean as is illustrated by the dashed blue lines in Figure~(6) where we plot the LOS scattering in the LRES
case for $z > 4.5$. The HRES case gives identical results with the FRES case and is ignored here.

Songaila and Cowie (2002) derived the following parametric fit
\begin{equation}
F(z,g) = 4.5 \, g^{-0.28} \, 
         \left ( {{1+z} \over {7}} \right )^{2.2} \,
         \exp \left ( -4.4 \, g^{-0.4} \, 
                      \left ( {{1+z} \over {7}} \right )^3 \right )
\end{equation}\smallskip\noindent
for the mean Ly$\alpha$\ transmission as a function of redshift which 
closely matches the $\Lambda$CDM calculations by Cen \& McDonald (2002) in the redshift range of z=6-4.
In the context of an uniformly ionized IGM, parameter g is the normalized ionization rate
(Mc Donald \& Miralda Escud\'e 2001).
\begin{equation}
g ~=~ \Gamma_{-12}~T_{4}^{0.75}~(\frac{\Omega_{m}}{0.35})^{0.5}~ \,
(\frac{\Omega_{b} h^{2}}{0.0325})^{-2} \,
\times (\frac{H_{o}}{65~Km~s^{-1}~Mpc^{-1}})
\end{equation}\smallskip\noindent
Following Songaila \& Cowie (2002) and Songaila (2004)  we assume a power law dependence of the form $g
\equiv g^{fit}=b_{1}~(\frac{(1+z)}{6})^{b_{2}}$ and calculate the pair of coefficients $b_{1}$ and $b_{2}$
that closely matches Equation~(3) to the computed optical depths in Figure~(6). Our calculation yields
$b_{1}=0.60$ and $b_{2}=-0.91$. The match between $MTF(z)$ and the analytic fit has an error of $\sim 4$\%
in the redshift interval z=4-6.4.  This tight fit is shown on the left panel of Figure~(7) (dashed line)
and is only valid in the range of $z=4-6.4$. The least squares fit is overplotted as a solid line and has
approximately the same margin of error ($4$\%) in the same redshift interval. However, the use of
Equation~(3) instead of Equation~(2) allows for the determination of the normalized ionization rate from
the synthetic spectra perspective. The solid line on the right panel of Figure~(7) is the normalized
ionization rate as inferred from the power law assumption in Equation~(3). The effect that the $1\sigma$
deviation of the scaling factor $b_{1}$, has on the ionization rate fit is shown as small dashed lines.  
The deviation to the power law exponent $b_{2}$, was not considered in this graph because the redshift
profile of Equation~(3) is very sensitive to that value.

The normalized ionization rate can also be computed directly from the simulation data through Equation~(4) if
we substitute for the gas temperature ($T_{4}$) the volume averaged temperature derived from our 
simulation data  
dumps. This rate is overplotted as the dashed-dot curve on the right panel of Figure~(7) and does not
have a power law profile. It is located however within the boundaries set by the $1\sigma$ deviation of the
scaling factor $b_{1}$ and has a mean absolute deviation from the power law fit ($g^{fit}$) of $\approx
20$\% in the redshift interval of interest (4-6.4).
An improvement to this rate can be sought (Appendix I) if we intuitively divide Equation~(4) by the ratio
$\frac{C_{HII}}{C_{B}}$ where $C_{HII}=\frac{<\rho_{HII}^{2}>}{<\rho_{HII}>^{2}}$ is 
the HII clumping factor and $C_{B}=<\delta^{2}>$ the baryon clumping factor.  
The adjustment was motivated by the fact that in a non-uniform IGM close to reionization the clumping
factor of ionized hydrogen (HII) is only approximately equal to the clumping factor of hydrogen (H). This
marginal effect is ignored in the derivation of Equation~(3). 
The latter accounts for a clumpy baryon distribution but not
for the relative difference between $C_{HII}$ and $C_{B}$. We compensate for that by adjusting the
normalized rate rather than the functional form of Equation~(3). The result is shown as a thick dashed
line on the right panel of Figure~(7). The mean absolute deviation between the raw simulation data and the
power law fit (spectra) is then improved from $\approx 20$\% to $\approx 12$\%. Nonetheless, the shape of
the input ionization rate still does not conform to that of a power law. 

The fits we considered to analytically represent the mean flux evolution at $z>4$ do not include the
last point at $z_{mean} = 6.52 \pm 0.08$ which is located above the fit-curves in Figure~(6) and
Figure~(7). In addition, the error of the fits to the flux data is improved if we also exclude the mean
flux at $z_{mean} = 6.36 \pm 0.08$. Those two points lie either inside or too close to the reionization
tail which in our setup is located at redshifts $z \geq 6.4$. To better relate observables to the
underlying physical properties we show on the top-left panel of Figure~(8) the profile of reionization, as
is traced by the mean baryon fraction in neutral hydrogen ($f_{HI}=<\frac{\rho_{HI}}{\rho_{B}}>$) between
z=5-7. The HI baryon fraction drops from $\approx 0.76$ at $z=6.8$ to $< 10^{-4}$ within $\Delta z = 0.4$.
We ramp up the ionization/heating rates from a tiny number ($10^{-30}$) at z=7 to the HM01 values at z=6.8
for numerical stability of the chemistry solver and therefore we do not trust computed species abundance
in that redshift range. At redshifts $z < 6.3$ $f_{HI}$ evolves smoothly with redshift which yields the
smooth evolution of the effective optical depth we measured in the synthetic spectra.

To quantify the reionization profile we fit the HI baryon fraction evolution between z=5-6.3 to a
linear-log parametrization, $log(f^{smooth}_{HI})~=~0.39( \pm 0.01)~(1+z)-6.96( \pm 0.03)$, and then
subtract the fit (extrapolated at $z>6.3$) from $log(f_{HI})$. The result measures the degree at which
$f_{HI}$ departs from the smooth evolution at $z>6.3$.  This 'reionization profile' is normalized by
its' maximum value and fitted to a step function ($F_{reion}$). The mean HI baryon fraction is then
recovered through $log(f_{HI}) = log(f^{smooth}_{HI}) + 4.06 \times (1-F_{reion})$. Each point along this
profile can be interpreted as the "Reionization Completion Parameter" (RCP). Reionization at 50\%
completion (RCP=0.5) sets the HI baryon fraction at about 100 times more that the one inferred by
extrapolating the smooth evolution from lower redshifts. The $F_{reion}$ profile shows that spectra
obtained between z=6.60-6.46 (the highest redshift interval in the effective optical depth plot) sample
the Ly$\alpha$ transmission when reionization is between the 20-98 \% completion level.

The MTF and variance evolution from our synthetic spectra are plotted in the high redshift interval
(z=5-7) on the bottom panels of Figure~(8). The mean transmitted flux (bottom-left) evolves with redshift
under the same power-law profile (solid-straight line) up to $z \approx 6.25$. It is then followed by an
order of magnitude decrease within $\delta z = 0.5$.  Overplotted are the margins of error to the MTF at
two confidence levels (CL) (bars: CL=68\% open: CL=90\%). The margins of error were determined by the
distribution of the LOS mean fluxes at each redshift bin and they systematically increase with redshift
for a fixed number of lines of sight. This is due to the increase with redshift of the LOS-variance which
is used to compute the margin of error. The LOS-variance dependence on redshift is expected because it is
effectively determined by the cosmic variance which increases as neutral hydrogen becomes more abundant.
Small values of cosmic variance will yield mean fluxes along a set of lines of sight that are closely
clustered about the LOS-averaged transmitted flux and therefore returning a small value for the
LOS-variance. On the other hand, mean fluxes computed along a set of lines of sight would be widely spread
(large LOS-variance) if the cosmic variance at that redshift has a large value.

The margin of error determines whether a particular mean transmitted flux value statistically deviates
from a smooth evolution or not. The MTF value in the redshift interval z=6.28-6.44 cannot be used to
conclude a deviation from a smooth evolution because the upper error margin (at both confidence levels)
includes the extrapolated curve from lower redshifts. That particular redshift interval should be
statistically excluded from being in the reionization tail even though it samples the last 1\% of it.
However, this conclusion is based upon the available number of lines of sight since the margin of error
scales as $(n_{los})^{-\frac{1}{2}}$. It is highly unlikely though that near future observations at high
redshift will exceed 75 Ly$\alpha$ transmission lines. If we consider fewer lines of sight then the
increased margin of error will only solidify our conclusion about the particular redshift interval.
Therefore, transmitted flux values recorded close to the end of reionization give no "statistical"
indication of sampling the tail despite the high probability that an individual transmitted flux value may
lie well beneath the smooth evolution curve.

The transmitted flux values measured in the redshift interval z=6.44-6.60 are located well within the
reionization tail. They in fact sample most of reionization's completion level curve. The mean transmitted
flux is 0.95 dex below the extrapolated power law of Equation~(2). The upper error bar at CL=90\% is also
below the power law curve by 0.55 dex which statistically places that redshift interval within
reionization. However, there is a non-zero probability that a single line of sight would yield a value for
the transmitted flux above $F^{fit}_{min} = exp(-\tau^{fit}_{max}(z))$ where $\tau^{fit}_{max} =
(2.1+0.16) (\frac{1+z}{6})^{4.16+0.02}$ from Equation~(2). That probability depends on the extent of the
redshift interval in which the LOS-spectra are sampled. The larger the extent the more low transmission
pixels are included from higher redshifts and therefore the probability of a high transmission pixel
becomes smaller. For $\Delta z \sim 0.16$ the probability that a line of sight would yield a mean flux
value more than $F^{fit}_{min}$ is $P=P(F_{LOS} > F^{fit}_{min}) \approx 1$\%.  If instead we resolve the
last redshift interval with four bins of $\Delta z = 0.04$ then P=[0.2,0.08,0.01,0.001] in each of the
intervals [6.44-6.48],[6.48-6.52],[6.52-6.56] \& [6.56-6.60] respectively. The mean redshift in each bin
corresponds to RCP values of RCP=[0.97,0.90,0.70,0.28] which suggests that the probability of a single
line of sight missing the last 10\% of the reionization epoch is more than 8\%. Although this number is
relatively small and model dependent it illustrates the importance of variance when synthesizing or
observing spectra in proximity to or within the reionization tail.

The right-bottom panel shows the evolution of the TMV-variance (solid line) and MLV-variance (dashed: FRES
dashed-dot: LRES). We note that the LRES variance is smaller when compared to the FRES case as expected.
The TMV-variance is the same in all cases since the mean flux is not affected by the spectral resolution.
The data were fitted to linear-log profiles for $z \leq 5.8$ which are overplotted in the figure. All
types of variance show a break from a linear-log profile for $z>5.8$ which is earlier (time moves
backward in this picture) than the corresponding break of the MTF from it's fit. However all cases remain
within $1\sigma$ from their lower-redshift fit curves up to $z=6.28$. In the following section, we examine
more closely the difference between the two types of variance we defined and their significance in mapping
reionization.

\subsection{Flux Variance Evolution} \label{fvar}

In section~(2) we defined the MTF-variance as the measure of dispersion of the LOS-mean flux values in
each redshift bin about the mean transmitted flux (MTF). In addition, we defined the MLV-variance as the
LOS-average of the variances along individual lines of sight. We plot in Figure~(9) the total MTF-variance
$TMV=NLOS \times \sigma_{MTF}^{2} $ and MLV-variance against redshift from z=2.5-6.6. The black triangles
show the TMV-data while the dashed and dot-dashed lines connect the MLV-data for the FRES and LRES cases
respectively. We find only a very small difference between FRES and HRES cases at $z<4.5$, therefore only
the last one is shown in that range. Overplotted as color-points are the variance values from the
individual lines of sight (green: FRES; blue: HRES; red: LRES) which are scattered about their respective
MLV lines. Across the redshift range z=2.5-5.8, a linear-log least squares fit in the form $log(Var)^{fit}
= A_{0}+A_{1}(1+z)$, matches our variance data from Figure~(9) with better than 5\% mean absolute
deviation. The constants $A_{0}$ and $A_{1}$ depend on the resolution and the type of variance and are
given in Table-I. The redshift point z=5.8 was selected visually because of an apparent break of the
TMV-data from a straight line there (solid line in Figure~(8)-Bottom Left panel). However, only data at $z
\geq 6.28$ have variance values $\| log(Var)-log(Var)^{fit} \| \geq \sigma_{A_{0}}+\sigma_{A_{1}}(1+z)$.
This redshift value matches the point of departure from a smooth profile of the MTF evolution shown in
Figure~(8) (Bottom Right panel) and corresponds to the redshift interval ($6.28 \leq z \leq 6.44$) which
samples the very last stages of reionization. It is however, the simultaneous significant break from a
smooth profile of both the mean flux and variance that a large enough margin of error is introduced to the
MTF to infer a possible continuation of a smooth evolution through that redshift interval.

The steep increase of the transmitted flux variance as we cross into the reionization phase is expected
because the variance at each redshift is equal to the total transmitted flux power (Tytler et al. 1997).
In turn the total transmitted flux power is proportional to the total mass power of neutral hydrogen which
decreases rapidly as reionization takes place.  Our computation volume is limited in that respect because
the largest length scale it simulates, (smallest wave-number) corresponds to the box size of
6.816$h^{-1}$ Mpc. Larger length scales can be sampled by wrapping a synthetic line of sight through the
simulated volume.  This is equivalent to assuming that the cosmic volume is an ensemble of such
volumes.  However, in doing so there is no additional gain in flux power. In this section we attempt a
description of the flux variance cosmic evolution which within the limitations of the present simulation
is a lower bound estimate.

A redshift profile with the functional form $Var = -1 + c_{o}(1+z)^{c_{1}}F^{c_{2}}$ ($c_{2}<0$) can be
inferred from theory (Appendix II) for the transmitted flux variance. This equation was derived by
computing the second moment of the local transmission over the volume distribution of densities
following Songaila and Cowie (2002) and it should generally be valid in the post-reionization era
(z=4-6.25) where Equation~(3) F(g,z) also fits our optical depth data. In the range of $\gamma =
1-\frac{5}{3}$ the constant $c_{1} \ll 1 ~ \Rightarrow ~(1+z)^{c_{1}} \approx 1$ (Appendix II). Using the
discrete LOS variance values in the FRES case from Figure~(9)  we can estimate that between $z_{1}=4$ to
$z_{2}=6.28$ the ratio $\frac{Var(z_{1})+1}{Var(z_{2})+1}$ scales between 16.6-25.7. This range compares
well with the one inferred from $\frac{Var(z_{1})+1}{Var(z_{2})+1} \approx (\frac{F_{1}}{F_{2}})^{c_{2}}=
16.6-22$ for $\gamma = 1$ and $\gamma= \frac{5}{3}$ respectively. In the last approximation, the values of
$F(z_{1,2})$ were obtained from the Equation~(3) fits to our data (Figure~(7)-Left panel).  The validity
of the analytic expression does not however suggest a deterministic dependence of the variance to the mean
transmitted flux but rather reflects that both quantities are correlated through their dependence 
on the mean IGM opacity.

Both the MTF-variance and the mean LOS-variance are measures of the "cosmic variance"  from two different
perspectives. The MTF-variance measures the normalized dispersion of a random sample of LOS-mean fluxes.
MLV in our setup measures the mean value of a random sample of normalized LOS-variances. Statistically the
mean value $\overline{x}$ of a sample of random values $x_{i}$ drawn from a larger population is a
point-estimate of the population mean $\mu$.  On the other hand, the variance of N random means
$\overline{x_{i}}$ underestimates the variance to the population mean by a factor N. Thus, we multiply the
computed MTF-variance by NLOS (number of lines of sight) to recover the total MTF variance (TMV) which is
an 'estimate' of the cosmic flux variance. Figure~(9) shows that for redshifts $z \leq 6.28$ both types of
variance have values within the scatter of the individual LOS data (points). The comparison also shows
that only the FRES/HRES MLV-data can be related to the total MTF-variance results which are not sensitive
to resolution.

Despite both variance types scaling in a similar manner with redshift, they are not equal because of the
different normalization and therefore a unique measurement of the cosmic variance of the transmitted
flux cannot be derived from either of them.
To directly compare one type of variance to the other, we are required to revert to the
standard (un-normalized) variance definition. The standard total MTF variance is equal to $\sigma_{M}^{2}
= \sigma_{MTF}^{2} \cdot (MTF)^{2} \cdot NLOS$. The standard mean LOS-variance is similarly equal to
$\sigma_{L}^{2} = \frac{1}{NLOS}~\sum_{j=1}^{NLOS} \sigma_{j}^{2} F_{j}^{2}$ where $F_{j}$ is the mean
flux along a line of sight j \footnote{ $\sigma_{L}^{2}$ is used in literature to describe the linear mass
variance. Here the notation stands for the weighted mean of the variances along lines of sight.}. In
theory, if the individual LOS-mean fluxes ($F_{j}$) are independent random measurements then
$\sigma_{M}^{2} = \sigma_{L}^{2}$ (Appendix III).
 
On the top left panel of Figure~(10) we plot $Log(\sigma_{M}/MTF)$ and $Log(\sigma_{L}/MTF)$ in the
redshift range z=2.5-6.6. The first quantity is TMV from Figure~(9) (triangles).  The second quantity is
the renormalized MLV-variance in the FRES case (dashed-line). 
Parametric fits to the data following a linear-log profile are provided in Table-I, 
which shows that the $\chi^{2}$ of the fits increases if we include values up to $z \leq 6.25$ compared
to the fits obtained from $z \leq 5.8$.   
Nonetheless, the curves only break away from a linear-log profile above the $1\sigma$ scatter of the fits  
upon crossing into the reionization phase in the redshift interval
[6.28-6.44]. The value for the variance increases by $\approx 2$dex between $z_{mean} =
6.2-6.5$. The decrease in the MTF (ignoring errors) is $\approx 1.4$dex in the same $z_{mean}$ range. This
difference in the degree of change of the two quantities (mean flux and flux variance) within the
reionization tail has a significant implication. The baryon mass in our computation box has a mean value
equal to the cosmic mean and therefore we do not expect the calculation of the MTF to be appreciably
biased by the finite volume. However, as discussed previously, our calculation of the flux variance in our
finite volume, yields at best a lower-bound estimate. This implies that a bigger computation volume or
line of sight observation, which would better sample the mass power spectrum at large scales, would in
turn yield a larger flux-variance and a bigger degree of change within the reionization tail. Therefore,
we can conclude that the difference in the degree of change within the reionization tail between the mean
flux and flux variance would be generally larger than the 0.6 dex we measured in this work. This suggests
that the flux variance exhibits a greater sensitivity to the evolution of the HI distribution than the
mean flux does and it could be used instead of the latter in mapping reionization.

The above conclusion is based upon the assumption that both types of variances we defined are estimates of
the cosmic variance of the transmitted flux in our synthetic Ly$\alpha$ spectra. 
However, the validity of such an assumption
depends on whether the two estimates agree with one another. The logarithmic y-axis on the top-left panel
of Figure~(10) might 'mask' any significant differences between the two quantities. We address this by
plotting the ratio between $\sigma^{2}_{M}$ and $\sigma^{2}_{L}$, each normalized to the square of the
mean transmitted flux, in the redshift range z=2.5-6.6 (Top-left panel of Figure~(10): squares).  For
comparison, we overplot (triangles) the ratio where we substitute in the denominator the mean LOS-variance
($\sigma^{2}_{MLV}$). The agreement between the normalized to MTF $\sigma^{2}_{M}$ and $\sigma^{2}_{L}$
generally holds well throughout the redshift interval. On the other hand, $\sigma^{2}_{MLV}$ agrees with
$\left ( \frac{\sigma_{M}}{MTF} \right )^{2}$ up to $z \approx 5.5$.  At redshifts $z>5.5$ the mean
LOS-variance consistently yields smaller values than the weighted by the ratio $(F_{LOS}/MTF)^{2}$ average
LOS-variance. The difference can be explained by the fact that $\sigma^{2}_{MLV}$ is a straightforward
arithmetic average of the individual normalized variances along each line of sight.  At redshifts close to
reionization high transmission regions become rare. Most lines of sight in fact do not include high
transmission regions in the last redshift interval (z=6.44-6.62). Therefore, at redshifts close to
reionization, the arithmetic mean of variances will be biased by the large number of small flux variance
spectra along lines of sight which do not include high transmission regions. On the other hand, the
weighted average LOS-variance is dominated by the few high transmission regions, as we will explain in the
Section~(5), and therefore it yields a larger value.
At redshifts $z < 5.5$ the mean LOS-variance and weighted average LOS-variance have approximately the same
ratio to the mean transmitted flux variance. This is not unexpected despite the different normalization.
Small values for the individual LOS-variances (at low to medium redshifts) suggest that the individual
LOS-mean fluxes are less scattered about the corresponding MTF value. Therefore, if $F_{j} \approx MTF$
(j=1,NLOS) then $(\sigma_{L}/MTF)^{2} \approx \sigma_{MLV}^{2}$.

Finally, it is important that one samples the cosmic Ly$\alpha$ transmission with a statistically adequate
number of lines of sight. Any random sample introduces a margin of error which becomes important to
minimize at high redshifts where the cosmic variance is large. Small number of lines of sight will retain
the skewness of the original population of transmission values but as long as the distribution is highly
peaked then the global average of mean fluxes can be trusted. For our 75 lines of sight, equal to the
number of mean fluxes per redshift interval, we compute the redshift evolution of the skewness and
kyrtosis of the mean flux sample. Following Croft et al. (1998) they are defined as $\gamma(z) =
\frac{<(x-1)^{3}>}{\sigma_{x}^{4}}$ and $\kappa(z) = \frac{<(x-1)^{4}>}{\sigma_{x}^{6}}$ respectively.  
In our notation, $x = \frac{F_{j}}{MTF}$ ($j=1,NLOS$), $\sigma_{x} \equiv \sigma_{MTF}$ and $<>$ denotes
an arithmetic averaging over all lines of sight. The results are shown in the lower two panels of
Figure~(10). We find that the distribution of line of sight mean fluxes per redshift interval has on the
average a skewness that fluctuates about zero with redshift and but is consistently positive at redshifts
close to reionization ($z>5.5$). The kyrtosis on the other hand consistently decreases with redshift which
shows that the distribution of LOS-mean fluxes becomes flatter. Therefore, if one requires an estimate of
the global average of the transmitted flux with a fixed margin of error, a progressively increasing with
redshift number of lines of sight is needed in order to compensate for the flattening of the distribution
of mean fluxes, which arises from the increasing cosmic variance.

On the left panel of Figure~(11) we plot the margin of error relative to the MTF as it scales with
redshift. At the 90\% confidence level (CL) the margin of error is $\sim 60$\% the MTF value at $z=5.5$
($\sim 40$\% at CL=68\%) and larger than 100\% at $z>6.4$ for both confidence levels.  This shows that 75
lines of sight undersamples the distribution of mean fluxes at high redshifts. In distributions with
significant skewness the margin of error is different for the two ends. The margin of error reported in
Figure~(11)  for those cases is the average margin of error between the two tail ends.  If we assume that
the LOS-mean flux measurements are normally distributed at all redshifts then we can estimate the LOS
number required to determine the margin of error within 10\% of the MTF. For our simulated cosmic volume
we would need 1200 lines of sight for the MTF to be computed within 10\% (CL=90\%)  at $z~\approx~5$. At
CL=68\% the required number of lines of sight drops to about 600 at $z~\approx~5$. The acquisition of a
few hundred high redshift Ly$\alpha$ observations is currently unattainable and therefore the large
margins of errors due to small size samples are unavoidable.

\section{Properties of the Flux Distribution}\label{cds}

The distribution of LOS-means at high redshifts is positively skewed because they are derived from the
positively skewed transmitted flux distributions along the individual lines of sight. Therefore extreme
LOS-mean fluxes at high redshifts are associated with rare high transmission regions. That in turn results
in a few "large" LOS-mean fluxes dominating the mean transmitted flux (MTF). The few high transmission
regions also dominate the variance. If $\sigma_{M}^{2}= \sigma_{L}^{2}$ then the MTF-variance is the
weighted sum of the individual LOS-variances ($\sigma_{j}^{2}$), $\sigma_{MTF}^{2} = (NLOS) \times
\sum_{j=1}^{NLOS}~\sigma_{j}^{2}~W_{j}^{2}$, where $W_{j}=\frac{F_{j}}{\sum_{j=1}^{NLOS}F_{j}}$. If there
exists a number of high transmission lines with LOS-mean fluxes $F_{k} \gg F_{j \neq k}$ then $W_{k} \gg
W_{j \neq k}$ and therefore the variance calculation would be biased by such lines of sight.

On the top panels of Figure~(12) we plot the transmitted flux (left panel)  and flux variance (right
panel) against redshift for $z>4.5$ while eliminating the statistical contribution of the rare high
transmission regions at large redshifts. We do that by substituting for $MTF(z)~\rightarrow~MODE(F_{j}(z))
\equiv MDF(z)$. $MODE(X_{j})$ selects the most frequent LOS-mean transmitted flux value. $F_{j}(z)$
denotes the mean flux of line of sight j, at redshift z.  The mean fluxes from our 75 lines of sight were
binned into 24 optical depth bins from the minimum LOS-mean flux value (converted to optical depth) to the
maximum one at each redshift. The number of lines of sight within each bin were then counted and the mode
was computed from the peak of the distribution. The variance was calculated by modifying the
$\sigma_{MTF}^{2}$ formula as follows. If $N_{mode}$ is the number of lines of sight with mean fluxes
within the bin of the mode and k is the array of indices the represents those LOS then $\sigma_{md}^{2}
\equiv N_{mode} \times \sum_{j=k}~\sigma_{j}^{2}~W_{j}^{2}$.  Our intent is to measure how the low
transmission lines of sight at high redshifts, which are more numerous and therefore have a higher
probability of being measured, affect mapping reionization.

Our calculations show that the high-z low transmission lines of sight (solid curve)  will indeed
underestimate the global average of the MTF value (dashed curve) by $\sim 1.6$dex (solid curve in lower
left panel)  during the last redshift interval (z=6.44-6.60), which samples the reionization tail.
However, the use of such lines of sight will not alter the instance where the flux redshift evolution
breaks from a smooth profile. In fact, the break is steeper when compared to the MTF evolution. Therefore,
a sample of low transmission lines of sight might offer insight to when the reionization is almost
complete but they will not offer a reliable measurement of the global average of the transmitted flux
during reionization. On the other hand, rare high transmission lines of sight at high redshifts will miss
the break of the MTF evolution at the end of reionization and infer the continuation of a smooth profile.
Such is the case of the cross points on the left panels of Figure~(12), where we plot the redshift
evolution of the mean transmitted flux based on lines of sight that have LOS-mean values $>10^{-3}$ in the
last redshift interval. Our conclusion is reinforced by the calculation of the variance for the low
transmission lines of sight which is indeed smaller for $z>6.0$ when compared to the global MTF variance
computed in the previous section (right panels).  The computed difference of $\sim 0.4$dex (in standard
deviation units)  in the last redshift interval may be considered small however it increases the
confidence in the mean transmitted flux value inferred by a sample of high-z low transmission lines of
sight.

Our differentiation between low and high transmission lines of sight was based upon examining LOS-mean
values computed from the arithmetic mean of pixel flux values across the redshift interval of our choice
($\sim 0.16$). Therefore, the same type of bias exists toward high transmission regions when computing the
flux average along a single line of sight. This effect depends on the size of the redshift interval. With
fixed redshift resolution, the smaller the size of the redshift bin at high redshifts the more susceptible
our mean value is to the presence of a high transmission region. If the majority of the transmitted flux
has values much less than the arithmetic mean then the size of the redshift bin should be increased to
give 'statistical power' to the low transmission segments within a line of sight.

On the left panel of Figure~(13) we plot the discrete flux distribution (DFD) in constant logarithmic bins
($\Delta logF=0.16$)  in the redshift interval z=5.95-6.45, in the LRES case only. The curve was computed
by averaging the number of pixels from all lines of sight in each bin and hence the presence of error bars
at large flux values.  The curve was then normalized to the total number of pixels.  For reference, we
overplotted the mean-flux (solid-line) and the upper-bound extreme of the Becker gap (shaded area). The
graph shows that low transmission regions have the largest probability of detection at high redshifts. We
can then compute from the graph, that flux values less than the upper-bound of the Becker-gap have a
probability of $\sim 80$\% of being recorded in our synthetic spectra. However, the mean flux computed in
the redshift interval and represented by the dashed line, lies in high transmission flux bins. In the
right panel of Figure~(13) we plot the mean transmitted flux in each of the flux bins of the DFD curve is
plotted against the mean flux in each bin.

In addition to the discrete flux distribution curve we also compute the Cumulative Distribution of
transmissions (Rauch et al. 1997), following Songaila and Cowie (2002). On the left panel of Figure~(14)
we plot the Cumulative LRES-Flux Distribution (CFD) for the redshift bins (from bottom up) (4.25-4.75),
(4.95-5.45),(5.45-5.95). In addition, we include the DFD redshift bin (5.95-6.45) which samples the
transmitted flux close to the reionization tail. The data points for the first three bins are extracted
sample values from the observed CDF-curves in Figure~(8) from Songaila and Cowie (2002).  For each
transmitted flux threshold (x-axis)  we average the fraction of the flux that lies below the threshold
value (y-axis)  from all lines of sight and compute the standard deviation.  The shaded areas are included
between the $\pm 2 \sigma$ curves.  There is a general agreement with the observed data in the low and
medium transmission region. We believe that the disagreement in the high transmission end of the x-axis is
due to the sensitivity of the CFD to the extrapolated continuum in the observed data.

The right panel of Figure~(14) is an x-axis logarithmic zooming into the CDF profile of the last redshift
bin. Similarly to the right panel of Figure~(13) we overplot the Becker-gap mean flux and upper-bound
extreme. The curve shows that $\sim 80$\% of all the flux pixels in the redshift interval z=5.95-6.45 are
below the Becker upper-bound extreme.

\section{Dark Gap Distribution}\label{dgp}
  
In the previous sections, we have determined that high transmission regions in the Ly$\alpha$ forest at
large redshifts become rare close to the epoch of reionization.  Therefore an alternate method to
analyzing the statistical properties of the transmitted flux is instead the distribution of dark gaps
(Croft 1998). By convention a transmission gap is defined as a contiguous region of the spectrum with
$\tau > 2.5$ over rest-frame wavelength intervals greater than $1 ~ \AA$. The gap distribution is then the
frequency distribution function of the gap wavelength width (GWW). Songaila \& Cowie (2002) obtained the
gap distribution in the Ly$\alpha$ region of the high redshift quasars BR 1202-0725 and SDSS 1044-0125 in
four redshift intervals (3.5-4.0), (4.0-4.5), (4.5-5.0) and (5.0-5.5) which showed a slow variation below
z=5.0 and a rapid increase in the number of gaps at $z>5$. In this section, we repeat their analysis using
our synthetic spectra samples adding two higher redshift intervals, (5.5-6.0) and (6.0-6.5).

In Figures~(15,16) we plot the gap width distribution (GWD) against the GWW in constant logarithmic bins
of size 0.25. The number of gaps per GWW-bin is averaged from all lines of sight and then normalized to
the redshift path, $\Delta X = \frac{2}{3} \Omega_{m}^{-0.5} [(1+z_{x})^{\frac{3}{2}} -
(1+z_{m})^{\frac{3}{2}}]$, where $z_{x}$ and $z_{m}$ are the upper and lower bounds of the redshift
interval. Overplotted in Figure~(15) are the observed data (diamonds) from Songaila \& Cowie (2002). The
vertical lines are the observational error bars. If no diamond-points are associated with a vertical line
then what is plotted is an upper bound estimate.  In order to compare with the observed redshift evolution
of the GWD, we consider contiguous gaps where the Ly$\beta$ optical depth also satisfies the dark-gap
threshold ($\tau_{Ly\beta} > 2.5$). The Ly$\beta$ optical depth is derived from the sum of the direct
Ly$\beta$ absorption and the Ly$\alpha$ absorption at the redshift $1+z_{\beta} \equiv
\frac{\lambda_{\beta}} {\lambda_{\alpha}} (1+z)$ (Songaila \& Cowie 2002).

\begin{equation}
\tau_{Ly\beta} = \frac{f_{Ly\beta} \lambda_{Ly\beta}}{f_{Ly\alpha} \lambda_{Ly\alpha}}
~\tau_{Ly\alpha} + \tau_{Ly\alpha}(z_{\beta})
\end{equation}\smallskip\noindent
where $\frac{f_{Ly\beta} \lambda_{Ly\beta}}{f_{Ly\alpha} \lambda_{Ly\alpha}} = 0.16$ is the
ratio of the product between the oscillator strength and the resonant scattering wavelength
for $Ly\alpha$ and $Ly\beta$.
In Figures~(15,16) the red symbols (triangles: FRES, crosses: HRES \& squares: LRES)  show the computed
GWD distributions. There is a general agreement with the observations in matching the redshift evolution
of the GWD. However, our calculations generally predict higher frequency values in the in the GWW range
$0.25 - 0.75$ (in log units) at z=3.5-5.5. The LRES data are below the corresponding full/high resolution
results for GWW values $ < 0.75$ and therefore tend to be in closer agreement with the observed points.
The difference between the high and low resolution can be explained as follows. The sharp wings of two
high transmission regions, separating a single gap in the HRES/FRES cases are smoothed out in the LRES
case and this results in a smaller GWW value for the gap. Small size gaps are more sensitive to the
convolution process because the spread of the gap is more influenced by the wings of the two bounding high
transmission regions. Therefore, a reduction in the computed GWW value for the small size gaps effectively
causes the translation of the FRES/HRES distribution leftward. However, due to the cutoff value of 1 \AA
~in the GWW size some gaps would be dropped in the LRES GWD calculation.

All computed GWD curves have power-law segments but also exhibit a power-law break below a redshift
dependent GWW value. When compared to the observed results in the bin z=4.5-5, which have a clear power
law distribution, the power-law portion of the computed GWD has a steeper slope. Our calculations
essentially mean that we detect more smaller size gaps than observed and most likely fewer larger size
gaps. We believe that this is due to our limited box size. Due to our fixed integration redshift step
($\delta z = 0.1$), each line of sight would cross the simulated volume a number of times before
progressing to the next redshift dump. Therefore, there is a non-zero probability that a number of
different lines of sight could register a resonant absorption feature from the same cosmic neighborhood.
Thus, an excess in the number of gaps of a particular size, when compared to the observed value, may be an
effect of oversampling. On the other hand, the size of our simulated volume does not adequate sample the
high end of the mass power spectrum. In massive objects, like large galaxies, the large abundance of
neutral hydrogen would register as large dark gap features in a transmission spectrum.  The absence of
such features in the computed GWD curve could be due to the lack of such objects in the volume or that,
our sample of lines of sight simply missed them.

Despite the disagreement on the exact shape of the GWD profile, our calculations reproduce in general the redshift
evolution of the gap distribution inferred by observations. The slow evolution toward larger gaps between z=3.5-5.0 
is followed by a rapid change in the GWD in observed and simulated data alike. Between z=5-5.5, both relay
an increase in the total number of gaps. Specifically, the number of gaps with GWW values $>0.75$ (in log units) 
exhibit a sharp increase which causes the gap distribution to flatten. Moreover, if we take into account the observed
upper limit error estimates then the observed power-law profile featured in the previous redshift bin
evolves to one similar to our calculation. However, we are compelled to note that the decrease of the
observed gap distribution in the smallest GWW bin ($1-10^{0.25}$ \AA) between z=5-5.5 does not occur in our 
synthetic spectra until $z \geq 5.5$. That may be due to comparing between results sampled and normalized 
by radically different number of lines of sight (2 versus 75) or that the reionization epoch inferred 
by the observed lines of sight occurs earlier than in our setup.

Figure~(16) shows that between z=5.5-6 the larger size gaps continue to grow in numbers while the smaller
ones increasingly disappear. This "gap size reshuffling" trend continues in the last redshift interval
which also includes gaps with sizes $> 100$ \AA. In addition, we plot in Figure~(17)  the total
number dark-gaps per redshift path which decreases at $z > 6$ after reaching a peak at $z \sim 5.5$. This
evolutionary trend of the GWD at high redshifts is more profoundly shown in the LRES case. A narrow "high
transmission" region separating two adjacent gaps in the full/high resolution cases maybe convolved to a
profile below the dark flux cutoff ($ < exp(-\tau_{cutoff})$) in the LRES case. This would manifest as a
merging between the individual gaps, which would disappear in the LRES sample, and the creation of a
single larger size gap. The LRES case simply illustrates more profoundly how the "gap size reshuffling"
occurs in all spectral resolution cases. As the redshift increases towards the reionization phase, not
only the frequency but also the amplitude of high transmission regions decreases.  Therefore, when the
transmitted flux in a previously "high transmission region" drops below the dark flux cutoff, the adjacent
gaps merge into a single one of a bigger size. This may explain why small size gaps rapidly disappear while
large gaps rapidly appear in the GWD between z=6-6.5.

The rate of the gap size increase is shown in Figure~(18) where we plot (left panel) the mean GWW against
the mean redshift in each interval.  We note the accelerated rate of increase at redshifts $ z \geq 5.5$.  
It is evident that the slope would have been steeper if the results were plotted against cosmic
time rather than redshift. The evolution progresses toward the reionization phase linearly with redshift
up to $z \approx 5.25$ followed by a steep power-law type increase. To infer a power-law profile we used
cubic interpolation between the data to acquire more points. Two least square fits are shown in
Figure~(18). A linear regression fit to the data at $z \leq 5.25$ yields the following parametric fit,
$<GWW> = a_{0} z + a_{1}$, where $a_{0} = -1.04(-1.67) \pm 0.34(0.48)$, $a_{1} = 0.70(0.87) \pm
0.08(0.11)$. The numbers in the parentheses refer to the low resolution data (dashed curves in Figure).  
For simplicity we treated the FRES and HRES cases as identical and used the full resolution data only. At
$z \geq 5.25$ a power-law fit in the form $<GWW> = (\frac{z}{z_{o}})^{b}$ yields $b = 8.874(10.089) \pm
1.095(1.282)$ and $z_{o} = 4.807^{+0.107}_{-0.134}(4.821^{+0.107}_{-0.136})$. We reported our fits in
high precision because the profiles are very sensitive to the exact values in a linear-linear plot. As
before, the parentheses refer to the LRES data. The normalized to the mean standard deviation
($(Variance)^{\frac{1}{2}}$) (right-panel) exhibits a similar evolutionary trend. A linear evolution up to
$z \sim 5.25$ is followed by a rapid increase under a strong power-law ($\frac{\sigma}{<GWW>} \propto
z^{4.64}$) for larger redshifts. As a result, the 1$\sigma$ deviation becomes comparable to the mean GWW
at about $z \sim 5.75$ which reflects the broadening of the GWD in the two high redshift intervals (Figure~(16)
Left-Panel). The extrapolation of our crude fits to z=6.9, which within $\delta z = 0.1$ corresponds to
the beginning of reionization in our setup, yields an estimate of the statistical mean value for the gap
wavelength width of $<GWW> = 37.2 \pm 90.7$ \AA ~(LRES fits). In comparison, the Ly$\alpha$ transmission
at z=6.9 in a redshift interval of $\delta z = 0.1$ corresponds to $\Delta \lambda = 121.6$ \AA ~which is
less than 1$\sigma$ from the mean gap width. In effect, we can characterize the entire spectral region at
that redshift as a "trough" and infer that the IGM is in effect neutral.

Our calculation begun by making an educated guess of the initial redshift and the profile of reionization
which allowed for the derivation of the IGM's global ionization properties and evolution.  Even though
synthesizing spectra at small RCP (Reionization Completion Parameter) values was impractical with our
method (Zhang et al. 1997), the extrapolation of fits to the redshift evolution of the mean gap width allowed 
for a consistent determination, within the statistical errors, of when reionization begun. Therefore, the evolution of the
mean gap width can be used as an alternative method of reionization mapping. The clear advantage of gap statistics
over using the mean transmitted flux at redshifts right after or during reionization, is that the MTF is
biased by high transmission regions, in real and redshift space alike, that become increasingly rare and
narrow as the redshift increases. Any conclusions drawn in that case will be affected by small number
statistics (too few lines of sight) and the large flux variance as we discussed in previous sections of
this work. On the other hand, as the dark gaps grow in size they include an increasingly larger portion of
the local optical depth along a line of sight and therefore volume fraction of the IGM, which is
significant if we are to make any claims about the IGM's global ionization state. Gaps also have the
advantage that they are insensitive to the exact flux values of the high transmission regions that bound
them. In addition, since the measurement of transmission values below the cutoff is more probable as we
approach the reionization phase, we expect that the line of sight variance will not be as important factor
as it was in examining the MTF properties. Nonetheless, a disadvantage of the gap size analysis we have
performed so far is that we can not directly infer quantitative properties of the underlying baryon distribution
inside the gaps since all opacity information is reduced down to a single optical depth value, the optical
depth threshold. For the remainder of this section, we investigate a possible correlation between the size
of a gap and some measure of the gap's optical depth. The motivation is that if such association exists it
is a measure of the coupling between dark gap statistics (size) and transmitted flux statistics
(amplitude).

One choice is the arithmetic "mean pixel optical depth" within each gap region.  In Figures~(19,20), we
scatter plot the mean Ly$\alpha$ against the mean Ly$\beta$ optical depth for each gap with wavelength
width spread larger than 1 \AA ~and for mean pixel optical depth greater than 2.5. The slope of the
straight line is equal to the ratio $\frac{f_{Ly \beta } \lambda_{Ly \beta }}{f_{Ly \alpha } \lambda_{Ly
\alpha }}= 0.16 $. The color bar on Figure~(19) shows the allocation of color for each pair of optical
depths based on their measured gap width. The points were sequentially plotted from the smallest to the
largest GWW value and therefore each color represents the largest gap value measured in the local\'e of a
mean optical depth pair $<(\tau_{Ly \beta}>, <\tau_{Ly \alpha}>)$. The solid colored regions on the graph
represent the "exclusion zones" where both the Ly$\alpha$ and Ly$\beta$ optical depths are smaller than
2.5. Since every pixel across a contiguous gap has optical depth above the cutoff threshold,
($\tau_{cutoff} \geq 2.5$), the average pixel optical depth in a gap is also above the same cutoff value.
This measure of a gap's optical depth would be heavily biased by the presence large pixel optical depth values and
is equivalent to associating the size of a gap to the opacity of the most underionized region(s) (largest
overdensities) it contains. 

The scatter of the data in Figures~(19,20) is due to the second term, $\tau_{\alpha}(z_{\beta})$, in
Equation~(5) which we used to compute the Ly$\beta$ optical depth. The data points are scattered upwards,
along the y-axis and off the constant slope line because of the contribution to $\tau_{\beta}(z)$ from
Ly$\beta$ photons redshifted to the Ly$\alpha$ frequency at a later redshift $z_{\beta}$. For
$\tau_{\alpha}(z)$ values between 15.625 - 2.5 the second term needs to contribute at least between 0 -
2.1 optical depth units respectively to $\tau_{\beta}(z)$ for the pixel to be part of a "dark gap".  For
$\tau_{\alpha}(z) \geq 15.625$ all associated Ly$\beta$ pixel optical depths are above the optical depth
cutoff.  Since the numbers of pixels with $\tau_{\alpha}(z)=15.625-2.5$ increases as the redshift
decreases, the identification of "dark gaps" fitting all the constrain parameters becomes increasingly
more dependent on the value of the $\tau_{\alpha}(z_{\beta})$ term.  On the other hand, as the redshift
decreases the average IGM opacity at $z_{\beta} < z$ is smaller ($\chi_{HI}(z_{\beta}) < \chi_{HI}(z)$)
while the baryon mass clumping is larger ($C_{\rho}(z_{beta}) > C_{\rho}(z)$). Consequently, it becomes
increasingly rare for the redshifted Ly$\beta$ photons to be resonantly scattered from a high opacity
region. On average, the contribution of the $\tau_{\alpha}(z_{\beta})$ term decreases with redshift and
that results in an increasing number of "off-slope dark gap" candidates being flagged out. The above
argument explains why the scatter of the averaged optical depth gap pairs decreases with decreasing
redshift.

We can measure the scatter off the constant slope line by computing the median ratio between the mean Ly$\beta$ and
Ly$\alpha$ gap optical depths at each redshift interval. It is difficult to gauge such relation from
Figures~(19,20) since in each local\'e of mean optical depths we can only visualize the largest gap size. Any
information of the scatter plot density is effectively masked out by the size of the plotting
symbols. For comparison purposes, we will also introduce an alternate method of measuring the
optical depth properties of a gap. That is the gap "effective optical depth" which is computed by the negative
natural logarithm of the transmitted flux averaged within the bounds of the dark region. Contrary to averaging
pixel optical depths which is sensitive to large values and therefore biased by the least ionized regions, this
approach relates a gap to the opacity of regions ionized the most within the gap's bounds.

The effective optical depth of a spectral region can be associated to a characteristic overdensity
value where $\tau_{eff}=\tau_{\Delta=1}~\Delta^{\beta}$. In this equation, $\tau_{\Delta=1} =
14g^{-1}~(\frac{1+z}{7})^{4.5}$ is the optical depth at the cosmic mean density and $\beta$ is a function of the
adiabatic index $\gamma$ (Appendix I,II). On the left panel of Figure~(21), we plot the redshift evolution of
$\tau_{\Delta=1}$, between $z=4-6.45$ for different choices of the normalized ionization rate, along with
$\tau_{eff}$ from Figure~(6). The latter is consistently below the $\tau_{\Delta=1}$ curves which indicates that it
corresponds to overdensities $\Delta < 1$. If we overplot about the effective optical depth the evolution of 
$\tau_{\Delta=0.36}$, for three adiabatic index values using the ionization rate inferred by the
Equation~(3) fit to our data ($g^{fit}$), we see that the three curves contain the redshift evolution of the
effective optical depth. Therefore, the latter can be associated to a characteristic overdensity $\Delta_{c}
\approx 0.36$. The exact value scales with redshift as shown on the right panel where we compute $\Delta_{c}$
from $\tau_{eff}=\tau_{\Delta=1}~\Delta_{c}^{\beta}$ in the redshift interval z=4-6.4. Irrespective of the normalized
ionization rate type or adiabatic index, which are listed on the figure, the characteristic overdensity
$\Delta_{c}$ associated with the effective optical depth decreases with increasing redshift. For
$\gamma=\frac{4}{3}$ and $g=g^{fit}$, $\Delta_{c}$ scales between 0.42 to 0.32 at z=4 and z=6.4 respectively.
In conclusion, the effective optical depth is biased by the highly ionized low overdensity regions.

The previous statement is based on properties of the entire spectrum regardless of an optical depth cutoff. 
Nonetheless, we can apply the method to the transmission fraction sampled 
by the dark gaps. Since that fraction increases as we approach
reionization, it is expected that the characteristic overdensity associated with the gap effective optical depth at high redshifts
to closely match the one inferred by all transmitted flux pixels. We first compute in Figure~(22) the redshift
evolution of the scatter between the mean optical depth and the effective optical depth for each dark gap in the
redshift intervals we considered. The two types are not correlated for mean optical depths larger than $ \bar{\tau}_{gap} \approx 15$ 
at $z \leq 6$ but are positively correlated at $z \geq 6$. They are also positively correlated 
for $\bar{\tau}_{gap} \leq 15$ at all redshifts. We then proceed to convert the scatter plots in Figure~(22) 
to the scatter plots in Figure~(23) between the characteristic overdensities
$\Delta_{c}$ estimated by $\Delta_{c} = (\frac{\tau}{\tau_{\Delta=1}})^{1/\beta}$ for $\gamma=4/3$ and
$\tau=\bar{\tau}_{gap}$ (mean optical depth) or $\tau=\tau_{eff}^{gap}$. The figure shows that the effective optical
depth is biased by smaller overdensities than the mean optical depth. In the redshift range approaching
reionization the characteristic overdensity inferred by the effective optical depth of the dark gaps is entirely in the
underdense ($\Delta < 1$) range. An overdensity average 
in the last redshift interval (6-6.5) yields $0.42 \pm 0.08$ which is comparable to $\approx 0.34$, the value derived
from Figure~(21) at $z_{bin}=6.25$ for the same $\gamma$ and choice of the normalized ionization rate. For comparison, 
the overdensity average inferred from the gap mean optical depth in the same redshift range is 
$1.28 \pm 0.78$ which statistically samples the cosmic mean density.

It is evident from the preceding analysis that the two gap optical depth definitions offer two
competitive perspectives on the properties of the underlying baryon distribution sampled 
within the dark regions 
and therefore together they yield a more complete description of the gap opacity evolution.
We can now plot on the left panel of Figure~(24) the evolution of the median ratio between the 
mean Ly$\beta$ and mean Ly$\alpha$ optical
depths from Figures~(19,20) (blue histogram). In addition, we also plot (red histogram) the median ratio if we
instead use the effective optical depth of a gap. As expected, the redshift evolution of both
histograms show a progressive reduction of the median ratio at smaller redshifts which
illustrates the decrease of the data-scatter off the constant slope line observed in 
Figures~(19,20). The asymptotic value of the ratio in the mean optical depth case 
approaches 0.16 by z=3.5 as predicted by Equation~(5).
Both computed ratios show that the Ly$\beta$ optical depth is an important selection 
bias in the flagging process of a low transmission region as a "dark gap". According to
Figure~(24), at redshifts following reionization completion, the mean (effective) gap Ly$\beta$ optical depth 
can be as much as $\sim 40$\% ($\sim 80$\%) that of the mean (effective) Ly$\alpha$ optical depth. 
A detailed examination of the redshift profiles in each case, which primarily depend on the
spectral slope and evolution of the UVB (Songaila \& Cowie 2002), is beyond the scope of this
work. We can however, estimate their relative scale averaged across the redshift
range. The evolution of the gap mean optical depth yields an average median offset
$\bar{\tau}_{Ly\beta}/\bar{\tau}_{Ly\alpha} - 0.16 \sim 0.14$ between z=3.5-6.5. In addition,
we can postulate that during the post-reionization smooth evolution of the IGM opacity, 
the offset in the effective optical depth ratio can be approximated by 
$\tau_{eff}^{\beta}/\tau_{eff}^{\alpha} - 0.16 \approx \phi$, 
where $\phi \equiv \tau_{eff}^{\alpha}(z_{\beta})/\tau_{eff}^{\alpha}(z)=
(\frac{1+z_{\beta}}{1+z})^{q} = (\frac{\lambda_{\beta}}{\lambda_{\alpha}})^{q}$.
Substituting for the power law exponent $q = 4.16$ from Equation~(2) we can then
estimate that $\tau_{eff}^{\beta}/\tau_{eff}^{\alpha} \sim 0.66$ and an average scale between
the two median rations of $0.66/0.3 \sim 2.2$ which is consistent with the 
median histograms in Figure~(24).

We have shown in Figure~(22) that in general, $\tau^{eff}_{gap} \leq \bar{\tau}_{gap}$ at all redshifts. This 
is also illustrated on the left panel of Figure~(24) where the median ratio of Ly$\beta$ to Ly$\alpha$ optical
depths is larger in the case of $\tau^{eff}_{gap}$. In scatter plot layouts
similar to Figures~(19,20), points corresponding to pairs ($\tau^{eff}_{\alpha},\tau^{eff}_{\beta}$) 
will be distributed in regions further to the left of the constant slope line when compared to $\bar{\tau}_{gap}$ pairs. 
However, as we shall shortly show, the Ly$\alpha$-Ly$\beta$ distributions for $\tau^{eff}_{gap}$ 
lack an important attribute that the $\bar{\tau}_{gap}$ distributions have. 
In Figure~(20), the scatter plots for the last two redshift intervals, gaps with sizes larger than 100 \AA
~(red color symbols), measured after $z=6$ in our setup, correspond to mean Ly$\alpha$ optical depths
larger than $\tau_{Ly \alpha} \approx 100$. This suggests a correlation between the gap size and the gap 
mean optical depth which is in fact illustrated by the way the data are plotted. The
sequential color plotting from smaller gap sizes to larger ones reveals that there is a trend in the color
scheme(size) to move to the right along the x-axis (Ly$\alpha$ mean optical depths) and upwards along the y-axis
(Ly$\beta$ mean optical depths), in other words toward larger mean optical depth values.
In order to measure this trend, we plot on the right panel of Figure~(24) 
the redshift evolution of the "Pearson Correlation Coefficient"
between the distributions of the gap Ly$\alpha$ mean optical depth ($\bar{\tau}_{gap}$) and wavelength width (GWW).
We note a weak but positive correlation persistent throughout the redshift range, with values between $r \sim
0.4 - 0.55$ (blue histogram). The same calculation between $\tau^{eff}_{gap}$ and GWW 
yields a positive but much weaker correlation ($r \sim 0.16$) in the last redshift interval only. 
For $z \leq 6$ the gap effective optical depth and size are uncorrelated.
This result is not a surprise because in Figure~(22) we showed that there is
little correlation between $\tau^{eff}_{gap}$ and $\bar{\tau}_{gap}$ at $z \leq 6$.
Finding a positive correlation between $\bar{\tau}_{gap}$ and gap size precludes
a positive correlation between $\tau^{eff}_{gap}$ and gap size. The two optical depth types
are largely uncorrelated because the mean optical depth is sensitive to a 
more extended range of baryon overdensities. As the dynamical range of overdensities becomes
smaller at large redshifts the correlation improves. 

Intuitively, the result on the right panel of Figure~(24) can be understood in terms 
of the opacity averaging procedure in each case. The mean optical depth method gives equal 
statistical weight at each local optical depth value along the gap. Therefore, the bigger 
the size of a gap the more extended the dynamical range of opacities sampled and consequently 
the direct mean being a "point-estimator" of all opacity data within the gap will reflect that.
Hence, the two quantities will be to some degree correlated irrespective of redshift.
On the other hand, the effective optical method gives equal statistical weight to the 
local flux. The optical depth estimate in this case is biased by a 
fraction of the gap, the regions of lowest opacity associated with the low 
end of the overdensity range sampled along the gap. 
Hence, the total size and the effective optical depth of a gap are
in principle uncorrelated. However, the approach of reionization in the 
last redshift interval shifts the overdensity bias in this method entirely 
in the underdense range, shown in Figure~(23), which dominates the volume filling factor.
Thus the segment of the gap to which the effective optical depth is sensitive to
rapidly increases to a size comparable to the gap width, hence
the sharp change from the uncorrelated profile at $z \leq 6$ to a weak but positive 
correlation at $z=6-6.5$.
A similar jump of the same magnitude is also evident in the blue histogram, 
however it is a more dramatic effect to traverse from completely uncorrelated data 
to some correlation at larger redshifts. 

The exact values of the correlation coefficients in each case will depend on the redshift
bin-size. The bin-size of $\Delta z = 0.5$, which was chosen for constructing
the gap width distribution, is most likely too large to showcase the 
transition to high correlation coefficients between size and optical depth 
as we cross into the reionization phase. Our values in the interval z=6-6.5 
from the histograms on the left panel of Figure~(24) are probably smooth out averages
between large values at $z \sim 6.5$ and small values at $z \sim 6$.

\section{Conclusions and Summary} \label{conc}

In this work we investigate properties of the Ly$\alpha$ transmission at redshifts during and after
reionization. We have performed a numerical simulation of the Ly$\alpha$ forest where ionization of the
primary species is due to a homogeneous UVB (Fig.1,2) which we switch on at a redshift that is consistent
with numerical and observational results which place the era of reionization at $z > 6.5$ (Fig.3). We
synthesized noiseless synthetic spectra along a fixed number of lines of sight through the simulated
volume and measured the Ly$\alpha$ flux transmission between z=6.6-2.5 (Samples shown in Fig.4). In
addition to the full pixel resolution spectra we obtain two more by convolving our spectra down to the
HIRES KeckI and the ESI resolution values.

Because of limits placed by our finite box size, we made no effort to explicitly match high redshift data,
rather study a realization of the Ly$\alpha$ forest consistent with the results by Songaila (2004), that
asserts a smooth profile for the effective optical depth beyond the Becker trough. The specific input
parameters for our calculation were chosen to match low and intermediate redshift data derived in previous
simulations (Jena et al. 2004). Our reionization profile is for all practical purposes artificial. It
serves the purpose of measuring the effect a reionization phase has on observable quantities by it's mere presence in the
high redshift universe. We use the Reionization Completion Parameter (RCP) as a dimensionless variable to
quantify the degree of reionization based on deviations from a smooth profile of the mean HI ionization
fraction. In doing so we introduce a visualization technique to probe the fraction of the reionization 
profile that our observable quantities sample (Fig.8).

By placing the beginning of reionization at z=7, we achieve full reionization by $z \approx 6.4$ and
derive a smooth power law profile for the effective optical depth from low redshifts up to the tail of
reionization (Fig.5,6). The smooth profile of our data between z=4-6.4 can also be fitted by the Songaila
\& Cowie (2002) analytic parametrization of the mean transmitted flux which was based on expanding the
uniform IGM optical depth formulation to a clumpy baryonic distribution (Fig.7). The inferred slope for
the normalized photoionization rate is shallower than the fit to the observed data however the inferred
photoionization rate is consistent within errors to the one determined by our input UVB and computed
temperature distribution (Fig.7).

As we enter the reionization tail, the mean transmitted flux (MTF) diverges away from the profile
inferred from earlier redshifts which is the standard method of asserting the beginning of the
reionization tail. However, we found that the large margins of error due to the inter-related effect of
the cosmic and line of sight variances can be detrimental in the degree of certainty behind claiming a
reionization era measurement at low RCP values, especially when high resolution spectra are used (Fig.8).
A measurement within the reionization tail can be statistically excluded if the margins of error to the
MTF include the extrapolated smooth profile from earlier redshifts.

In Section~(4.3) we studied the evolution of the flux variance and found that the variance to the MTF is
inherently related to the mean line of sight of variance (Fig.10,11). Each line of sight variance is a
measure of the cosmic variance along that direction, which increases as the IGM becomes more neutral.
However, the increased variance along a line of sight introduces an increasing with redshift uncertainty
in measuring the mean flux value and that error is carried over to the MTF computation, which is the
average over all line of sight mean fluxes. In the end, the ability to determine a highly certain value
for the MTF at high redshifts is inadvertently degraded as is indicated by the small kyrtosis value of the
LOS-mean flux distribution within the reionization tail (Fig.11). The only way to statistically measure
the MTF value with a relatively small margin of error is to use an extraordinary number of lines of sight
which is highly unlikely to be obtained observationally (Fig.11). Therefore, we conclude that the
uncertainty in the mean transmitted flux is unavoidable. The problem is compounded by the use of a mean
flux analysis at high redshifts.

The mean transmitted flux is a quantity that is biased by high transmission regions and the contribution
of low transmission or dark segments of the spectrum is diluted away by the exponentiation of the local
optical depth. This is an important issue at high redshifts where such high transmission regions become rare.
However, their presence along a line of sight will result in a mean flux calculation which will closely
resemble the magnitude of the high transmission region (or few high transmission regions). This is the
basis of the distinct probability of having to contend with a few large LOS-mean flux values within our
sample of random lines of sight. Their subsequent contribution to the MTF will also skew the results
toward a larger value when compared to the result obtained if such high transmission LOS were to be excluded
(Fig.12). Because that skewed value enters the calculation of the MTF variance, the deduced margins of
errors will also be large and that finally will yield a picture that is statistically close to a smooth
profile at a redshift when we are clearly within the reionization tail.

A much clearer association between the properties of the transmitted flux and reionization is derived when
instead of the MTF we use the mode (most frequent value) of the LOS mean fluxes (Fig.12). In doing so, we
effectively exclude rare high transmission measurements at high redshifts. Unfortunately, what is more
probable or less probable in the high redshift universe does not directly translate to the ability of
observations to do measurements of the transmitted flux. In fact, due to noise and the selection bias of
sky objects, there is a higher probability to observe a high transmission line of sight rather than a
low transmission one which are more frequent but harder to measure. In the end, our conclusion is that
observing a simultaneous steepening of the redshift profiles of the MTF and the flux variance
(irrespective of definition, type and spectral resolution) relative to an extrapolated smooth evolution
solidly points to the time when reionization terminates. However, the measurement of global properties
of the IGM and the UVB during reionization depends on the profile of reionization itself which is 
difficult to derive because of the intrinsic high degree of uncertainty of the data. Another way
to look at our result is that each line of sight appears to correspond or effectively is an attempt to
measure it's own unique reionization profile along it's redshift path. The difficulty lies in ascertaining
a single global reionization profile even if one is provided. We have come to this conclusion even when a
uniform reionization UVB was used as an input. However, the assumed homogeneity of reionization through
the global average of the IGM opacity is a stretchy assumption. Even if the UVB is uniform, the network of
chemical reactions "operates" on an inhomogeneous baryonic distribution. The resulting opacity
distribution is non-uniform and depending on the scale at which we view it, inhomogeneous as well.

In Section~(5) we also examined the properties of the flux distribution at high redshifts (5.45-6.45)
(both discrete and cumulative) which shows that most of the transmitted flux ($\sim 80$ \%)  lies at
values below $\approx 6 \times 10^{-3}$ or the upper limit of the Becker (2001) trough.  In contrast, the
mean transmitted flux at these redshifts lies in the top 20 \% of the flux distribution (Fig.13,14). This
suggests that the mean flux at high redshifts reflects the opacity properties of a portion (volume
fraction) of the IGM which becomes smaller as we approach reionization. Therefore, even though Becker-type
troughs dominate at high redshifts, a mean transmitted flux analysis fails to register their
contribution.  Our conclusion is that the traditional method of measuring properties of the Ly$\alpha$
forest through the MTF is deficient and needs to be supplemented by an analysis that directly involves low
transmission regions.

In Section~(6), we studied the distribution of "dark gap" sizes and it's evolutionary properties which
becomes relevant at high redshifts because an increasingly larger fraction of the IGM is sampled within
them.  When compared to the observed data of dark gap sizes our results differ marginally in the exact
shape and extent of the distribution (Fig.15,16). That maybe due to the fact that the observed data we
considered consisted of only two lines of sight, when in this work we took into consideration all of our
75 lines of sight. Nonetheless, the redshift evolution towards reionization, from small to large
redshifts, is consistent with the one deduced from observations. The results show an accelerated
"creation" rate of larger size dark gaps at high redshifts (Fig.15). The inclusion of two higher redshift
intervals, which we did not have any observed data to compare to, reinforced the previous conclusion
(Fig.16). We suggested a mechanism of gap merging as a possible explanation of the occurrence of larger
gap sizes as we approach reionization and measured the redshift evolution of the average gap width
(Fig.18). Similarly to the mean transmitted flux analysis, the mean gap width and the error introduced by
the increasing with redshift spread of the gap distribution steepen as the profile approaches
reionization redshifts. Crude fits of the computed results to a power law functional form and
extrapolation to higher redshifts yields a mean gap size that with the error is comparable to the size of
the transmission spectrum. The redshift at which the above occurs coincides with the beginning of
reionization in our setup Therefore, we conclude that such an extrapolation of the mean gap width to high
redshifts, offers an appealing method in pinpointing the time at which reionization begun.

Finally, we found that there is positive correlation between the gap size and the mean optical depth of
the pixels it contains (Fig.19,20,24). Such correlation couples the size of dark regions to the underlying
opacity distribution. Therefore, we conclude that by combining the analysis of the mean transmitted flux
and dark gap sizes one can infer the properties of the entire IGM structure sampled by a line of sight
through the cosmic volume.

\section{Acknowledgments} 

We would like to thank Tridivesh Jena in providing up-to-date cosmological parameters for our simulations
that fit low redshift Ly$\alpha$ observations. Simulations were performed and analyzed at the San Diego
Super Computer Center (SDSC) and at the University of Illinois National Center for SuperComputing
Applications (NCSA). This research was supported by the National Science Foundation under grant
AST-0307690.

\section{Appendix I}

Standard calculation of an analytic approximation of the Ly$\alpha$ optical depth 
begins with the HI photo-ionization equilibrium condition that
relates the number density of HI to the photo-ionization rate ($\Gamma_{HI}$), the recombination
coefficient $\alpha (T)$, a function of the gas temperature, and the baryon density of HII under the
assumption of the electron abundance being determined by hydrogen ionization only:

\begin{equation}
n_{HI} = \frac{\alpha (T)}{\Gamma} n_{e} n_{HII} \approx
\frac{\alpha (T)}{\Gamma} (n_{HII})^{2} \rightarrow \,
n_{HI} = \frac{\alpha (T)}{\Gamma} n_{HII}^{2} = \frac{\alpha (T)}{m_{p}^{2} \Gamma}
\rho_{HII}^{2}        
\end{equation}\smallskip\noindent
  
In a uniform IGM after reionization $\rho_{HII} \approx \rho_{H} = \frac{\rho_{H}}{\rho_{B}} \rho_{B} =
Y_{H} \rho_{B}$, where $Y_{H}$ is the cosmic hydrogen abundance and $\rho_{B} = \overline{\rho_{B}}$ is
the baryon cosmic mean. Integrating the optical depth along a line of sight from redshift z to the present
yields the uniform Ly$\alpha$ optical depth expression $\tau_{u} = 14 g^{-1} (\frac{1+z}{7})^{4.5}$, where
g is the normalized photo-ionization rate given in Equation~(3). A non-uniform IGM approximation can be
obtained by assuming that $\rho_{HII} \approx \rho_{H}$ and integrating the rms value
$<\rho_{H}^{2}>$ of the 3D hydrogen distribution along a line of sight. Since
$<\rho_{H}^{2}> = Y_{H}^{2} <\rho_{B}^{2}> = Y_{H}^{2} \overline{\rho_{B}}^{2} C_{B}$, where $C_{B} =
<\delta_{B}^{2}>$ is the baryon clumping factor, the non-uniform IGM Ly$\alpha$ optical depth is then
$\tau_{Ly\alpha} = \tau_{u} C_{B}$. The condition $C_{B} = 1$ recovers the uniform expression. An
expansion can be obtained (Hui \& Gnedin 1997, Croft 1998, McDonald \&
Miralda-Escud\'e 2001) by assuming a power law dependence of the local optical depth to the local
overdensity $\tau_{\Delta} = \tau_{u} \Delta^{\beta}$ where $\beta = 2 - 0.75(\gamma-1)$. The mean
transmitted flux, Equation~(2), can then be computed (Songaila \& Cowie) by integrating $F=\int
P(\Delta)~exp(-\tau_{\Delta})~d \Delta$ where $P(\Delta)$ is the baryon distribution function
(Miralda-Escud\'e, Haehnelt \& Rees 2000).

The basis for the above approximations is the assumption $\rho_{HII} \approx \rho_{H}$. According to the
top-left panel of Figure~(8) this is good global approximation even during the late stages of reionization
($\geq 90$\%) where the volume averaged neutral fraction decreases from $\sim 10^{-3.5}$ at $z
\geq 6.5$. However, the volume averaged neutral fraction of hydrogen is biased toward low overdensities
which dominate the volume filling factor. At large overdensities, the approximation will not be as valid
and therefore there will be a discrepancy between the distribution of HII in respect to the underlying
baryons. In the context of homogeneous reionization that discrepancy is less significant when compared to
inhomogeneous reionization with self-shielding but it's effect can not be ignored at times close to
reionization. In our simulation the ratio of HII to Baryon clumping factors scales between 0.5 at $z=6.5$
and 0.95 by $z=6.2$. Because the previous optical depth functional forms do not account for this
"chemical" effect we do not allow the approximation $<\rho_{HII}^{2}> \approx <\rho_{H}^{2}>$ but still
use $<\rho_{HII}> \approx <\rho_{H}>$.
 
\begin{equation}
<n_{HI}> = \frac{\alpha (T)}{m_{p}^{2} \Gamma} \frac{<\rho_{HII}^{2}>}{<\rho_{B}^{2}>}
<\rho_{B}^{2}> = \frac{\alpha (T)}{m_{p}^{2} \Gamma} 
(\frac{<\rho_{HII}>}{<\rho_{B}>})^{2} <\rho_{B}^{2}> C_{HII}/C_{B} 
\end{equation}\smallskip\noindent
where the ratio $(\frac{<\rho_{HII}>}{<\rho_{B}>})^{2}$ can be approximated by $Y_{H}^{2}$.
Therefore, the previous equation can be written in the form
$<n_{HI}> = \frac{\alpha (T)}{m_{p}^{2} \Gamma} Y_{H}^{2} C_{B} \overline{\rho_{B}} 
C_{HII}/C_{B} \rightarrow \tau = \tau_{u} C_{B} \frac{C_{HII}}{C_{B}}$. 
The argument we make is that if the FGPA equation was the basis for expanding
$\tau = \tau_{u} C_{B} \rightarrow \tau_{\Delta} = \tau_{u} \Delta^{\beta}$
then to account for the difference in the clumping factor between the baryon and HII distributions 
at redshifts close to reionization we assume that Equation~(2)
applies under the expansion
$\tau = \tau_{u} C_{B} \frac{C_{HII}}{C_{B}} \rightarrow \tau_{u}^{c} \Delta^{\beta}$,
where $\tau_{u}^{c} = 14 g^{-1} \frac{C_{HII}}{C_{B}} (\frac{1+z}{7})^{4.5}$.
The ratio of the clumping factors is then effectively absorbed in the normalized
ionization rate.    

\section{Appendix II}

The mean transmitted flux, Equation~(2), was computed in Songaila \& Cowie (2002) 
from the integral
\begin{equation}
<F>=\int P(\Delta)~exp(-\tau_{\Delta})~d \Delta \approx
\int A \Delta^{-b}d \Delta~exp(-\tau_{u} \Delta^{\beta} + \frac{\Delta^{-\frac{4}{3}}}
{8\delta_{o}^{2}/9})
\end{equation}\smallskip\noindent
where $\tau_{\Delta}=\tau_{u} \Delta^{\beta}$, $\tau_{u}$ is the optical 
depth in a uniform IGM, $\beta = 2-0.75(\gamma-1)$
and $\delta_{o} = 7.61(1+z)^{-1}$. $P(\Delta)$ is the functional dependency of the volume
density distribution (Miralda-Escud\'e et al. 2000) on overdensity $\Delta$ and $b \approx 2.5$.
The integration via the method of steepest descents (Songaila \& Cowie 2002)
yields a general functional form for the mean transmitted flux:
\begin{equation}
<F> = A (\frac{4\pi}{3\beta+4})^{0.5} \delta_{o} \Delta_{o}^{5/3-b} \,
\times exp[- (\frac{3}{2\beta}+\frac{9}{8}) \Delta_{o}^{-4/3} \delta_{o}^{-2}] 
\end{equation}\smallskip\noindent
where $\Delta_{o}=(\frac{1}{2 \tau_{u} \beta \delta_{o}^{2}})
^{\frac{1}{\beta+4/3}}$ 
is the value of $\Delta$ at which the exponent in Equation~(8) sharply peaks. 
The function F(g,z) is then derived through the dependence of $\tau_{u}$ on the normalized 
ionization rate g, $\tau_{u} = 14g^{-1} (\frac{1+z}{7})^{4.5}$.
The mean square value of the transmitted flux is similarly expressed by integrating
$exp(-2 \times \tau_{\Delta})$ instead of $exp(-\tau_{\Delta})$ over the volume 
density distribution function. We can view the multiplication factor of 2 as a modifier of
a constant ($\tau_{u}$) in
the integration and therefore we simply rewrite Equation~(9) with 
$\hat{\Delta}_{o}= \Delta_{o} (\frac{1}{2})^{1/(\beta+4/3)}$ instead of $\Delta_{o}$.
\begin{eqnarray}
<F^{2}>=\int A \Delta^{-b}~d \Delta~exp(-2\tau_{u} \Delta^{\beta} + 
\frac{\Delta^{-\frac{4}{3}}} {8\delta_{o}^{2}/9}) \approx
A (\frac{4\pi}{3\beta+4})^{0.5} \delta_{o} \hat{\Delta}_{o}^{5/3-b} \\
\times exp[- (\frac{3}{2\beta}+\frac{9}{8}) 
\hat{\Delta}_{o}^{-4/3} \delta_{o}^{-2}]
\end{eqnarray}\smallskip\noindent
Substituting for the functional forms of $<F^2>$ \& $<F>^{2}$ in the variance definition,
$Var = \frac{<F^{2}>}{<F>^{2}}-1$, we get 
$Var =-1+ [A (\frac{4\pi}{3\beta+4})^{0.5} \delta_{o}]^{-1}
(\frac{1}{2})^{(5/3-b)/(\beta+4/3)}~\Delta_{o}^{b-5/3}~
exp[ (\frac{3}{2\beta}+\frac{9}{8}) \delta_{o}^{-2} \Delta_{o}^{-4/3}
(2-2^{1/(1+3\beta/4)}) ]$. If we set $q_{o} = (\frac{1}{2})^{\frac{5/3-b}{\beta+4/3}}$
and $q_{1} = (\frac{1}{2})^{\frac{3\beta/4}{1+3\beta/4}} < 1$ then the previous
equation becomes: 
\begin{equation}
Var = -1 + q_{o} [A~(\frac{4\pi}{3\beta+4})^{0.5}~\delta_{o}~\Delta_{o}^{5/3-b}~
exp(-(\frac{3}{2\beta}+\frac{9}{8}) \delta_{o}^{-2} \Delta_{o}^{-4/3}~2(1-q_{1}))]^{-1}
\end{equation}\smallskip\noindent
where the quantity $2(1-q_{1}=0.85-0.77$ between an isothermal 
($\gamma = 1$, $\beta = 2$) and an adiabatic ($\gamma = 5/3$, $\beta = 1.5$) equation of state
respectively. We can simplify Equation~(10) if we substitute in the expression 
for the mean transmitted flux to get:
\begin{equation}
Var = -1 + q_{o} [A~(\frac{4\pi}{3\beta+4})^{0.5}~\delta_{o}~\Delta_{o}^{5/3-b}]^{1-2q_{1}}~
F^{2(q_{1}-1)} = \,
-1+c_{o} (1+z)^{c_{1}} F^{c_{2}}
\end{equation}\smallskip\noindent
where $c_{o} > 0 $ \& $c_{2}=2(q_{1}-1) < 0$. The term $(1+z)^{c_{1}}$ is derived from
the power-law dependence of the $\delta_{o} \propto (1+z)^{-1}$ \& 
$\Delta_{o} \propto g^{-0.3} \times (\frac{1+z}{7})^{-0.75}$ (Songaila \& Cowie 2002) 
in addition to a power-law assumption for the normalized ionization rate, 
$g=b_{o}(1+z)^{b_{1}}$. The exponent then becomes $c_{1} = (1-2q_{1})
[(5/3-b)(0.3b_{1}-0.75)-1]$. Substituting for $b \approx 2.5$ \& $b_{1} = -0.91$,
the power law exponent inferred from our data in the redshift range $z=4-6.4$, 
we get $c_{1}=0.021-0.033$ for $\gamma = 1-\frac{5}{3}$. As a result, the term
$(1+z)^{c_{1}} \approx 1$ and therefore the variance depends primarily on the
value of the mean transmitted flux in a cosmic environment. 

\section{Appendix III}

Assume a random sample (lines of sight) of N means $X_{i}$ with measured
standard (un-normalized) variance $\sigma_{i}^{2}$. The standard variance of the mean 
$\overline{X}=\frac{1}{N} \sum_{i} X_{i}$ is then equal to
\begin{equation}
Var(\overline{X}) = Var(\frac{1}{N} \sum_{i} X_{i} ) = \frac{1}{N^{2}} \sum_{i} Var(X_{i}) \,
= \frac{1}{N^{2}} \sum_{i} \sigma_{i}^{2} = \frac{1}{N} <\sigma_{i}^{2}>  
\end{equation}\smallskip\noindent
where $<>$ denotes the average over the variance sample. If the variances along all lines 
of sight are equal to a single value (the cosmic flux variance, $\sigma_{cf}$, at redshift z)
then $\frac{1}{N} <\sigma_{i}^{2}> = \sigma_{cf}^{2}$ and therefore 
$Var(\overline{X}) = \frac{\sigma_{cf}^{2}}{N}$. 
In our case $Var{\overline{X}} \equiv \frac{\sigma_{M}^{2}}{N}$ which suggests
$\sigma_{M}^{2} = \sigma_{cf}^{2}$. In addition, we defined $\sigma_{L}^{2} = <\sigma_{i}^{2}>$ 
and therefore $\sigma_{L} \equiv \sigma_{M}$.

\begin{deluxetable}{lccc}
\tablewidth{450pt}
\tablenum{1}
\tablecaption{Regression constants in $Log(Variance^{\frac{1}{2}})=A(0)+A(1)~(1+z)$ for Figure~(9)-
first four rows- and Figure~(10)-last three rows. RNMLV-VAR refers to the renormalized - to the
MTF - mean LOS-variance: ($\sigma_{L}/MTF)^{2}$.
\label{tbl:1}}
\tablehead{
\colhead{Redshift Range \& Type} & \colhead{$\chi^{2}$}
& \colhead{$A(0) \pm \sigma_{A(0)}$} & \colhead{$A(1) \pm \sigma_{A(1)}$}  }
\startdata
[2.5-5.8]  TOTAL-MTF-VAR  & $3.8~10^{-2}$ &  $-1.627 \pm 0.050$ & $0.308 \pm 0.010$ \nl
[2.5-5.8]  FRES-MLV-VAR   & $2.8~10^{-3}$ &  $-1.459 \pm 0.013$ & $0.277 \pm 0.003$ \nl
[4.5-5.8]  LRES-MLV-VAR   & $1.8~10^{-4}$ &  $-1.589 \pm 0.026$ & $0.286 \pm 0.004$ \nl 
[2.5-4.5]  HRES-MLV-VAR   & $6.3~10^{-4}$ &  $-1.430 \pm 0.016$ & $0.273 \pm 0.004$ \nl
[2.5-5.8]  FRES-RNMLV-VAR & $2.4~10^{-3}$ &  $-1.452 \pm 0.013$ & $0.274 \pm 0.002$ \nl
[2.5-6.25] FRES-RNMLV-VAR & $2.4~10^{-2}$ &  $-1.532 \pm 0.032$ & $0.291 \pm 0.006$ \nl
[2.5-6.25] TOTAL-MTF-VAR  & $5.4~10^{-2}$ &  $-1.678 \pm 0.048$ & $0.319 \pm 0.009$ \nl
\enddata
\end{deluxetable}

\newpage
%

\begin{figure}
\epsfig{figure=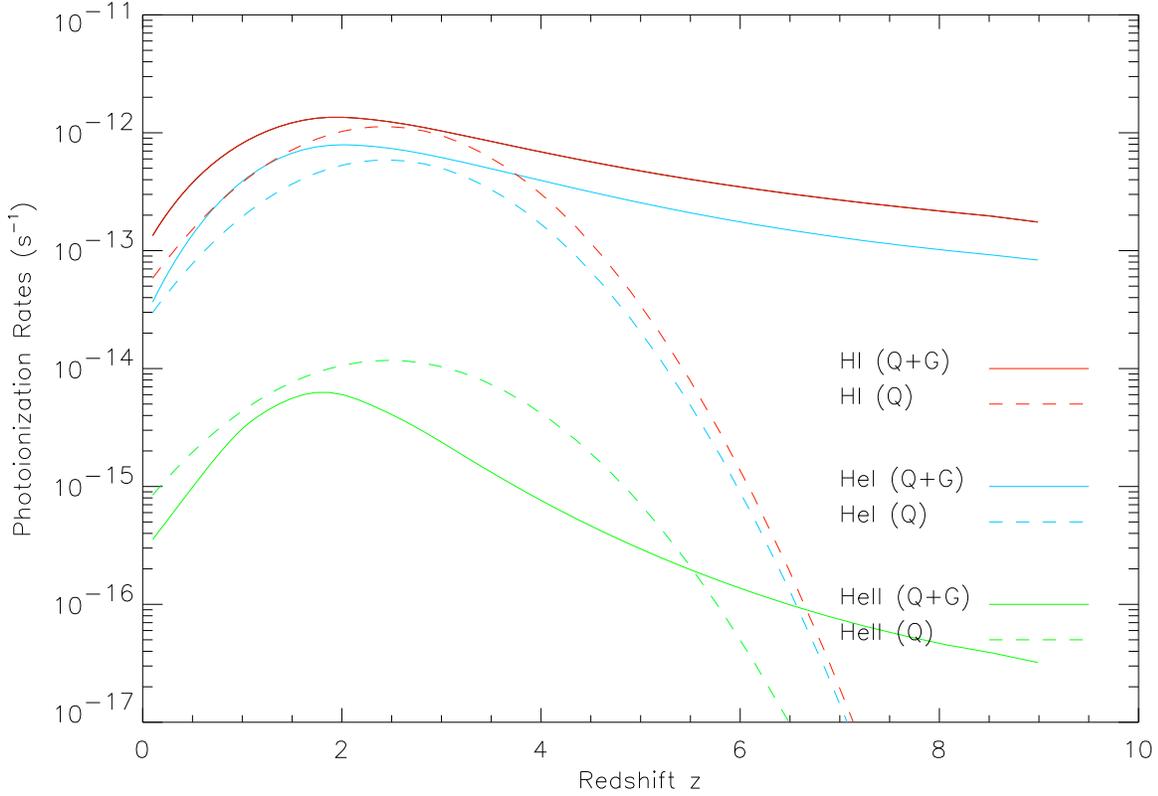,angle=0,width=6.0in}
\vspace{6pt}
\figurenum{1}
\caption{
Fits to the redshift evolution of the photo-ionization rate for the
three primary IGM species (HI,HeI \& HeII)  derived from the soft spectrum calculation
by Haardt \& Madau (2001)  (solid line) for our $\Lambda$CDM cosmology. The softness
of the spectrum ($\frac{\Gamma^{HI}}{\Gamma^{HeII}} ~ > ~ 100 $) is due to the UV
production in galaxies.  The larger number of galaxies in the high redshift Universe
(compared with QSOs) makes them the principal ionizing sources of neutral hydrogen.
The dashed line shows the rates computed due only to quasar contributions. The falling
QSO number counts in the high redshift epoch ($z > 4$) results in a steep drop in the
ionization rate.
}
\label{fig1}
\addtolength{\baselineskip}{10pt}
\end{figure}

\newpage
%
%
\begin{figure}
\epsfig{figure=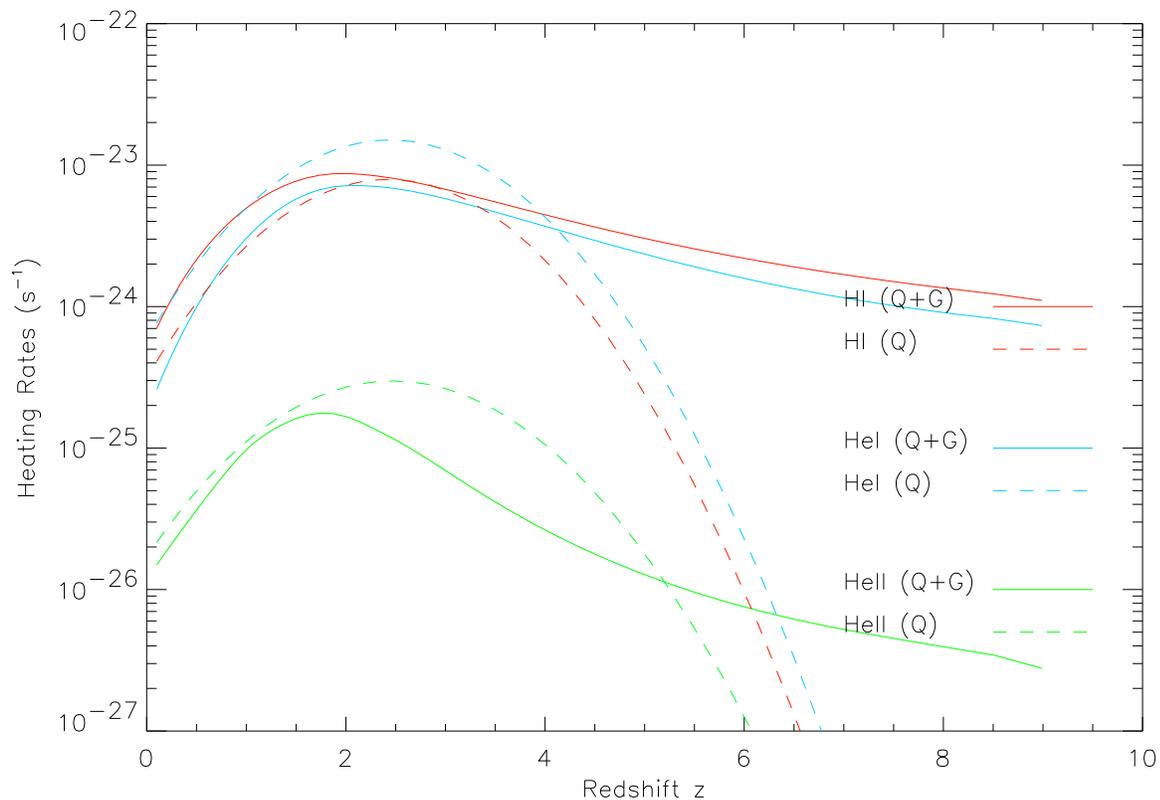,angle=0,width=6.0in}
\vspace{6pt}
\figurenum{2}
\caption{
As in Fig. 1 where instead of the photo-ionization rate per baryon we show the
photo-heating rates. The increased rates in the HM01(Q+G) case yield higher IGM
temperatures for $z > 4$ (compared to the QSO only case) which in turn results in a
grater thermal broadening of the Ly$\alpha$ lines. The increased thermal broadening
enhances the effects of line-blending and the creation of dark gaps in the transmitted
flux.
}
\label{fig2}
\addtolength{\baselineskip}{10pt}
\end{figure}

\newpage
%
%
\begin{figure}
\epsfig{figure=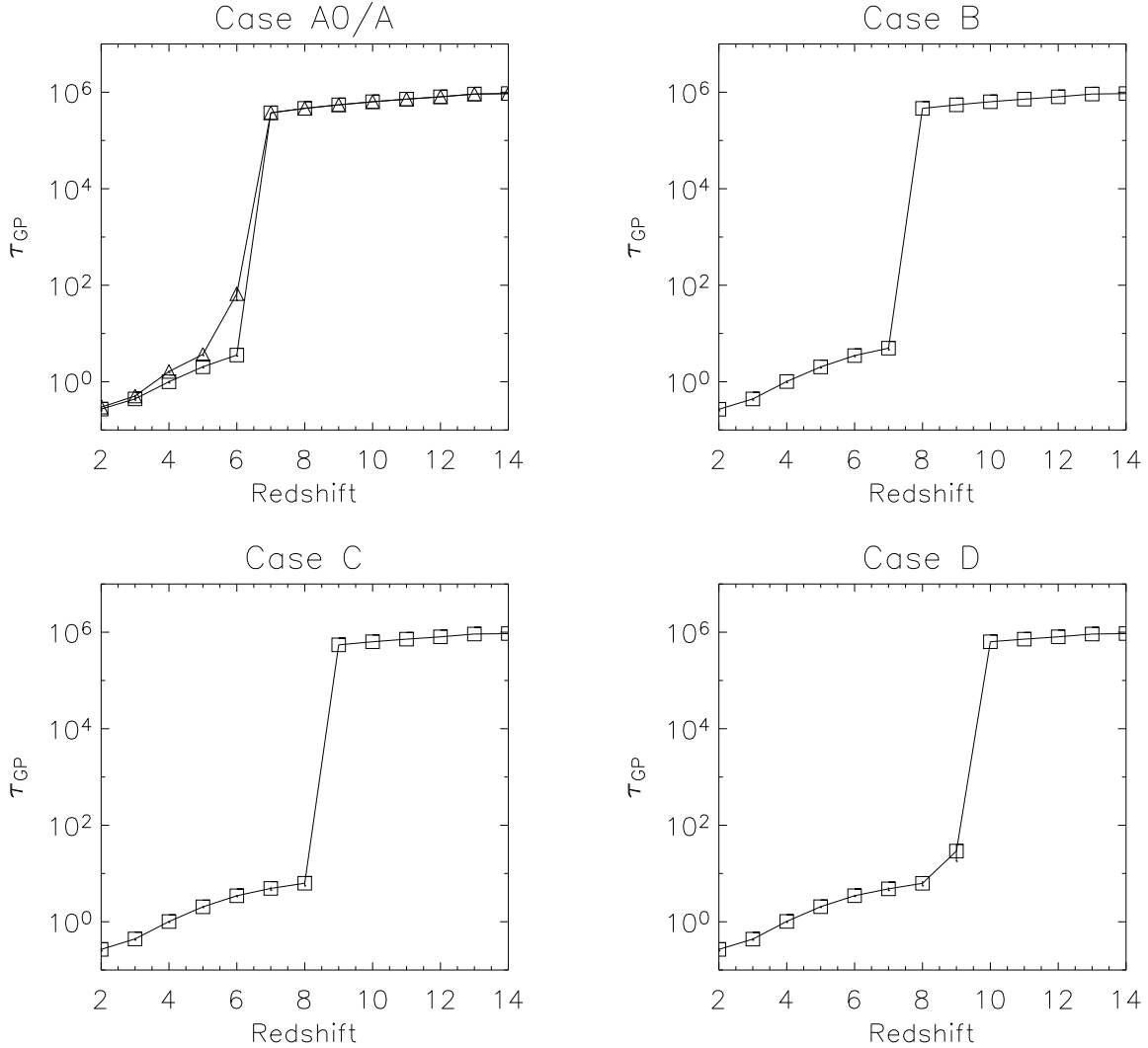,angle=0,width=6.0in}
\vspace{6pt}
\figurenum{3}
\caption{
The redshift evolution of the Gunn-Peterson optical depth is plotted for four cases:
Case A0 refers to the quasar-only HM96 UV flux spectrum (open triangles). 
Case A makes use of the HM(2001)
HM01 UV spectrum where $z_{on}=6.5$ (open squares). Cases B,C \& D are the same as Case A but with
$z_{on}$=7.0,8.0 \& 9.0 respectively. The choice of the $z_{on}$ does not effect the
evolution of optical depth in the redshift range of interest ($z\leq 6.5$).
}
\label{fig3}
\addtolength{\baselineskip}{10pt}
\end{figure}

\newpage

%
\begin{figure}
\epsfig{figure=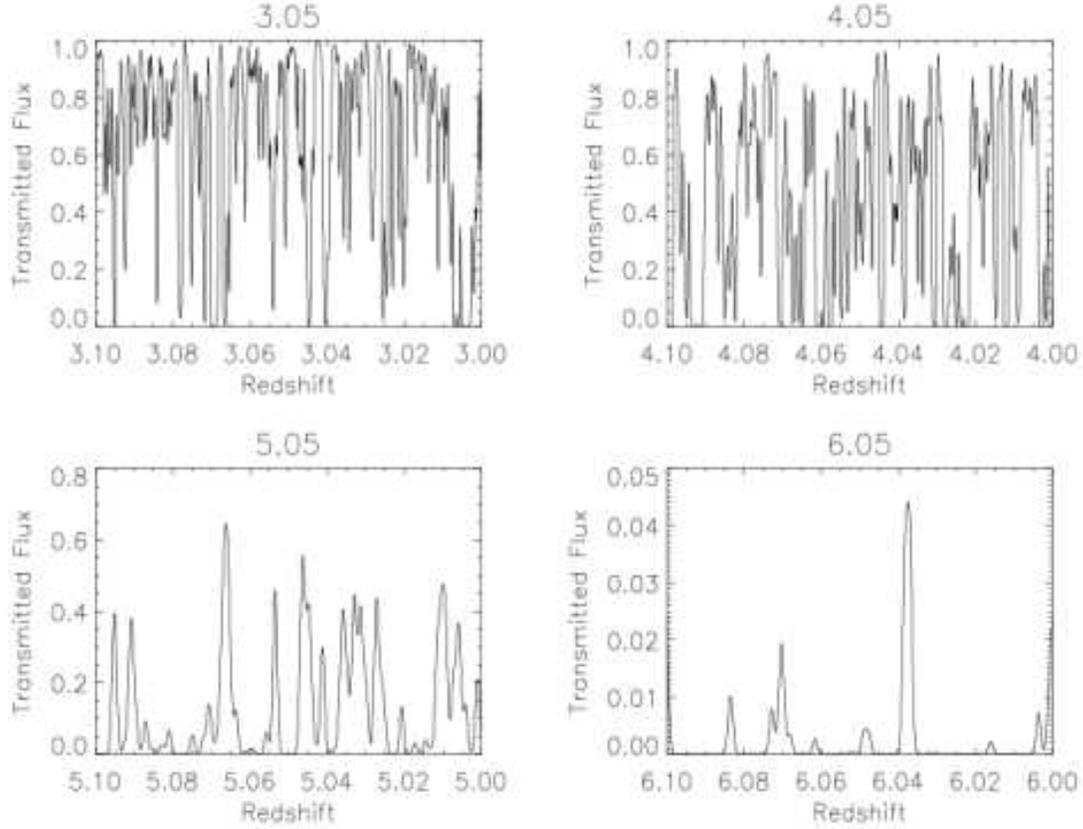,angle=0,width=6.0in,height=4.5in}
\vspace{6pt}
\figurenum{4}
\caption{
Extracted segments from a synthetic spectrum along a \emph{single} line of sight in
bins of $\Delta z=0.1$. The title in each panel is the mean redshift value in the
extracted range.  We have convolved our spectrum with a gaussian at the spectral
resolution of R=36,000 for the top two panels and R=5,300 for the bottom two panels in order
to emulate the observed HIRES KeckI data and ESI data at these resolutions
respectively. As the line of sight is cast through the time-series of the simulation
data-dumps it samples different regions of the computation volume. Therefore each
panel depicts the local Ly$\alpha$ absorption from a 
different cosmic neighborhood. 
}
\label{fig4}
\addtolength{\baselineskip}{10pt}
\end{figure}

\newpage

%
%
\begin{figure}
\epsfig{figure=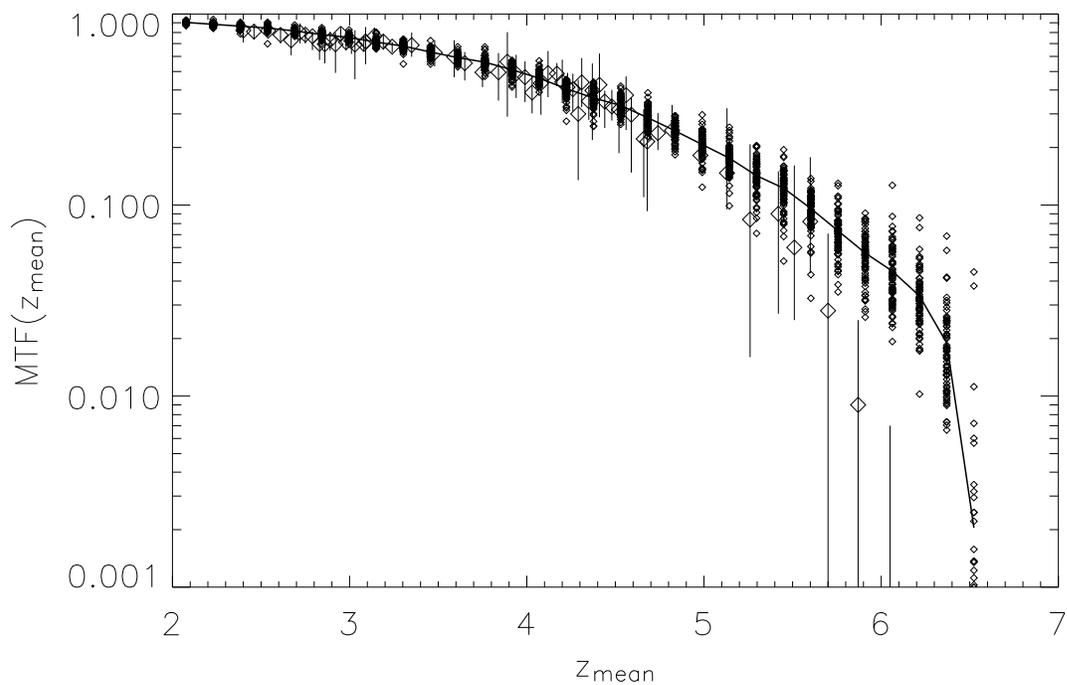,angle=0,width=6.0in,height=4.0in}
\vspace{6pt}
\figurenum{5}
\caption{
Mean Transmitted Flux of the Ly$\alpha$\ forest as a function of redshift as predicted
by our simulated data.
The solid line marks the averaged flux in 30 redshift bins of $\Delta
z=0.153$\ ($\Delta \lambda \approx  186~{\rm \AA}$)  from all the 75 lines of sight.
The small diamonds represent the individual mean fluxes from each line of sight.
Overplotted are the ESI samples for $z > 4.5$ and the HIRES
KeckI samples for $z<4.5$ from Songaila (2004) (large diamonds).
}
\label{fig5}
\addtolength{\baselineskip}{10pt}
\end{figure}

\newpage
%
%
\begin{figure}
\epsfig{figure=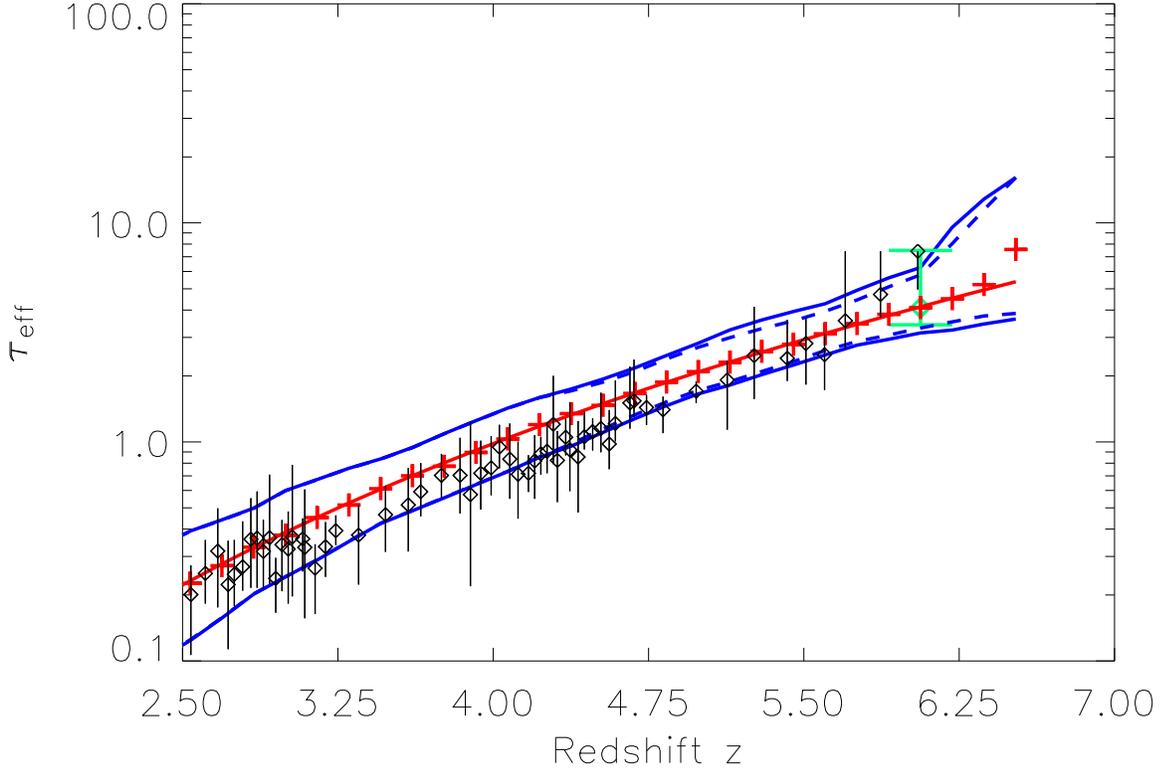,angle=0,width=6.0in,height=4.0in}
\vspace{6pt}
\figurenum{6}
\caption{
As in Fig. 5 where the mean transmitted Ly$\alpha$ flux (red crosses) is converted to
optical depth.  The observed flux lower limit (upper error bars in the optical depth
plot) is set to 0.0015.  The blue lines above and below the effective optical depth
represent the LOS-averaged extremes in each redshift interval (solid: FRES \& HRES; dashed: LRES).
The solid red line is a power law fit to the effective optical depth. The Becker gap (the highest redshift
diamond) is within $2.5\sigma_{LRES}$ (green error bar) from the MTF at $z_{mean}=6.06
\pm 0.08$.  The power law fit $\tau_{eff} = 2.1^{+0.12}_{-0.11}(\frac{1+z}{6})^{4.16
\pm 0.02}$ does not include the optical depth at the last redshift bin ($z_{mean}=6.52
\pm 0.08$).
}
\label{fig6}
\addtolength{\baselineskip}{10pt}
\end{figure}

\newpage
%
%
\begin{figure}
\epsfig{figure=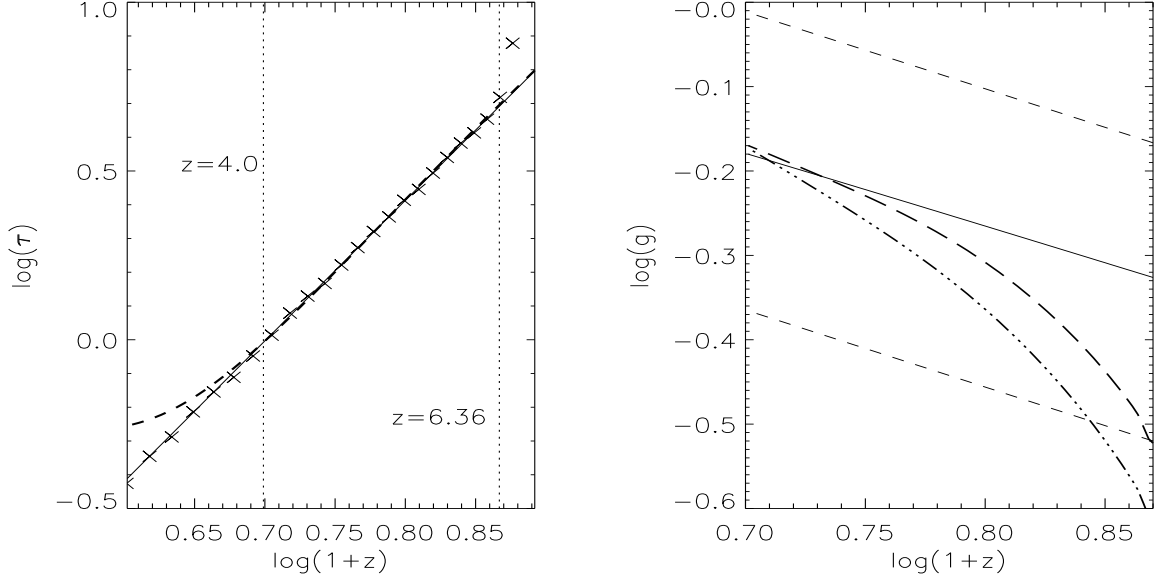,angle=0,width=6.0in,height=3.0in}
\vspace{6pt}
\figurenum{7}
\caption{
On the left panel we plot the effective optical depth with respect to redshift focusing on the
redshift interval 4-6.4. The crosses and the solid line are the effective optical depth 
and the power law fit from Figure~(6). The dashed line is the fit to the data using Equation~(3). The
fit is good to within 4\% error. \newline
On the right panel, we plot the normalized ionization rate versus
redshift in the same interval as on the left panel. The solid line is inferred from Equation~(3) under
the assumption of g having a power law dependence ($g=b_{1}~(1+z)^{b_{2}}$). 
The small dashed lines were computed from the 1$\sigma$ error
to the scaling factor in the power law ($b_{1}$). Direct application of simulation data in Equation~(4) 
yields the dashed-dot line. The long-dashed line adjusts the previous result by the ratio of
the HII clumping factor over the baryon clumping factor. 
}
\label{fig7}
\addtolength{\baselineskip}{10pt}
\end{figure}

\newpage
%
%
\begin{figure}
\epsfig{figure=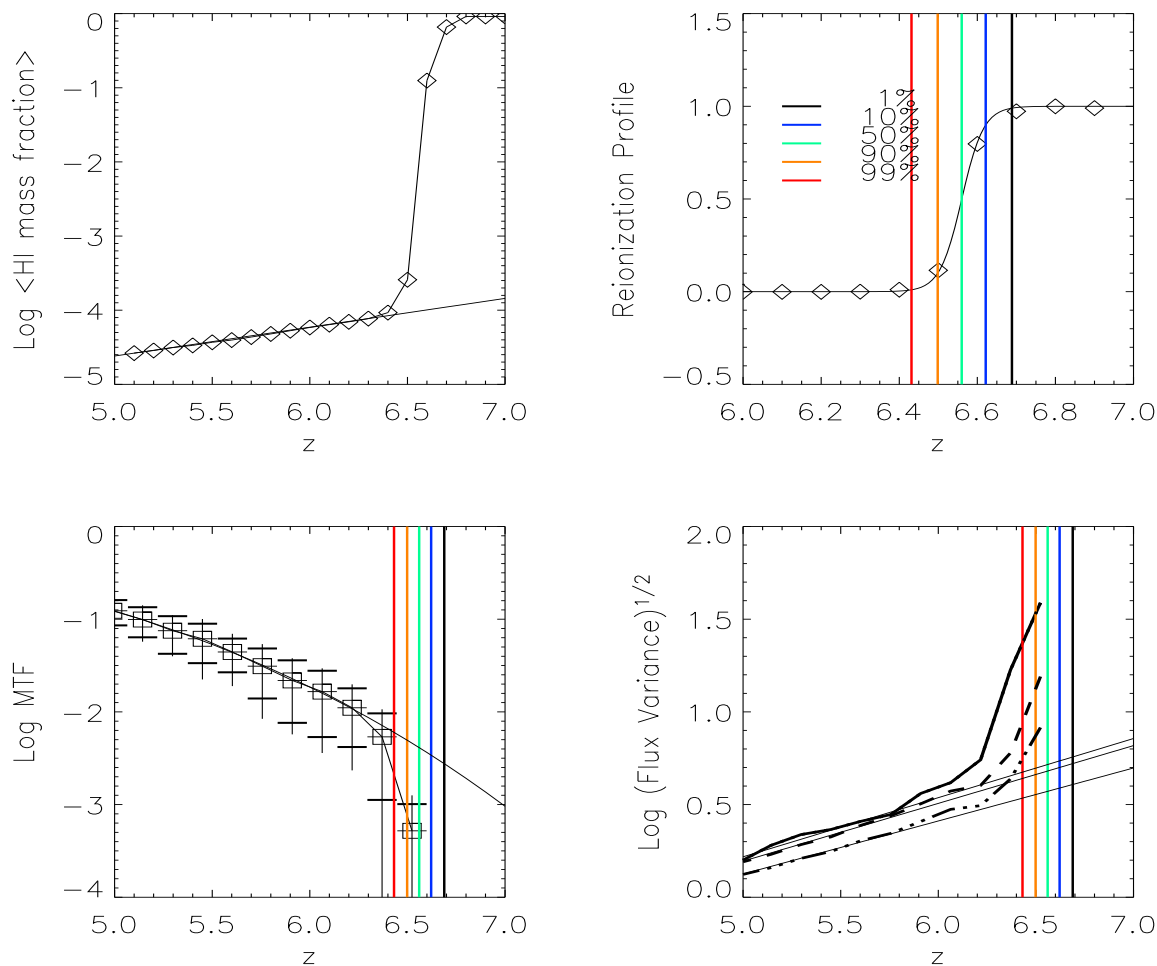,angle=0,width=6.0in,height=5.0in}
\vspace{6pt}
\figurenum{8}
\caption{
On the top panels, we show the profile of reionization as it is traced by the mean
baryon fraction in neutral hydrogen (top-left) between z=5-7. The re-ionization
profile (top-right) is fitted to an analytical function and the redshifts of
re-ionization percentage of completion are color-coded. Reionization is 99\% complete
by z=6.43 (red line) compared to 1\% completion at z=6.68 ($\Delta z_{reion} = 0.25$).
The MTF \& variance evolution in the same redshift range are plotted on the bottom
panels. The Log(MTF) ($\propto \tau_{eff}$) (bottom-left) 
evolves under the same power-law redshift profile (solid-straight line) 
up to $z \approx 6.25$. It is then followed by a 1-dex decrease within $\delta z = 0.5$.
Overplotted are the margins of error to the MTF (bars: CL=68\% open: CL=90\%). The
right-bottom panel shows the evolution of the total MTF-variance (TVAR: solid line) and
the Mean LOS-variance (MLV: dashed=FRES, dashed-dot=LRES). The straight lines are linear-log
fits to the data using redshifts $z \leq 5.8$. 
The MLV-variance breaks away from a linear-Log profile at $z \approx 6.25$ as it enters
the extreme end of the reionization tail.  A break in the TVAR curve at $z \geq 5.8$ is
not statistically important until $z \geq 6.25$.
}
\label{fig8}
\addtolength{\baselineskip}{10pt}
\end{figure}

\newpage

%
%
\begin{figure}
\epsfig{figure=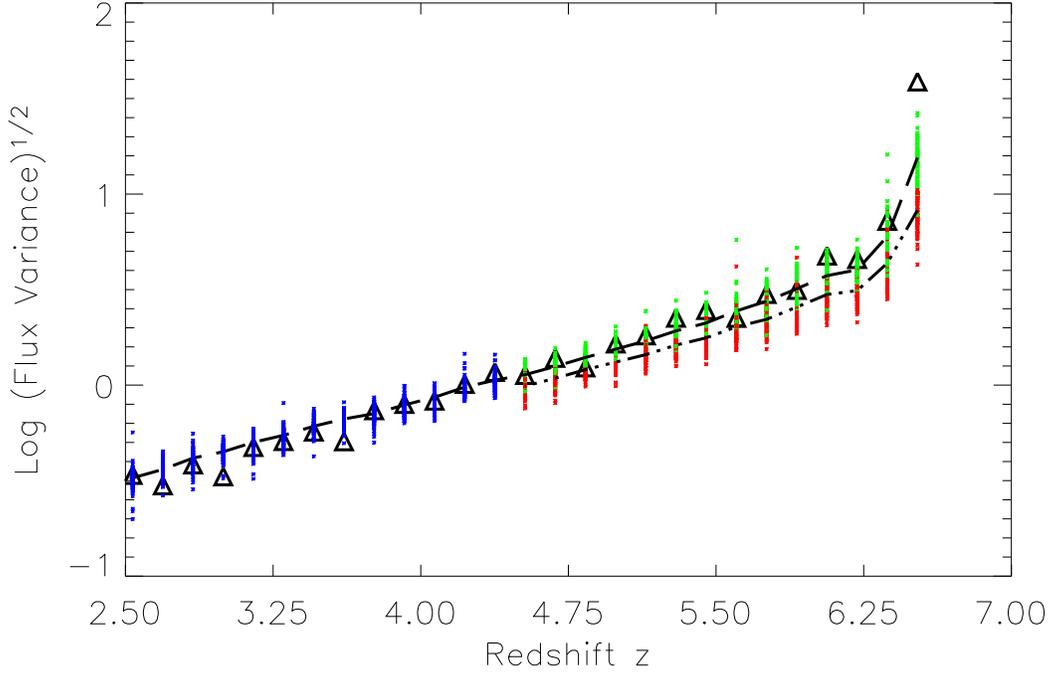,angle=0,width=6.0in,height=4.0in}
\vspace{6pt}
\figurenum{9}
\caption{
Redshift evolution of the Ly$\alpha$\ transmission variance as a function of redshift for
the complete sample of our synthetic spectra. The black triangles show
the total variance of the mean transmitted flux (MTF). 
The dashed and dash-dot lines show the MLV-variance data in each resolution case
which cross the distribution of individual LOS-variance values shown as colored-points 
(green/blue for FRES/HRES, red for LRES). 
The profiles evolve with redshift under a linear-log scaling law which fits the data up to $z \sim 5.8$
(Table-I). Between $z=5.8-6.25$ the data vary within the standard deviation of the linear-log fit.
At $z \geq 6.25$ all variance types steeply rise as we cross into the reionization epoch.
}
\label{fig9}
\addtolength{\baselineskip}{10pt}
\end{figure}

\newpage

%
\begin{figure}
\epsfig{figure=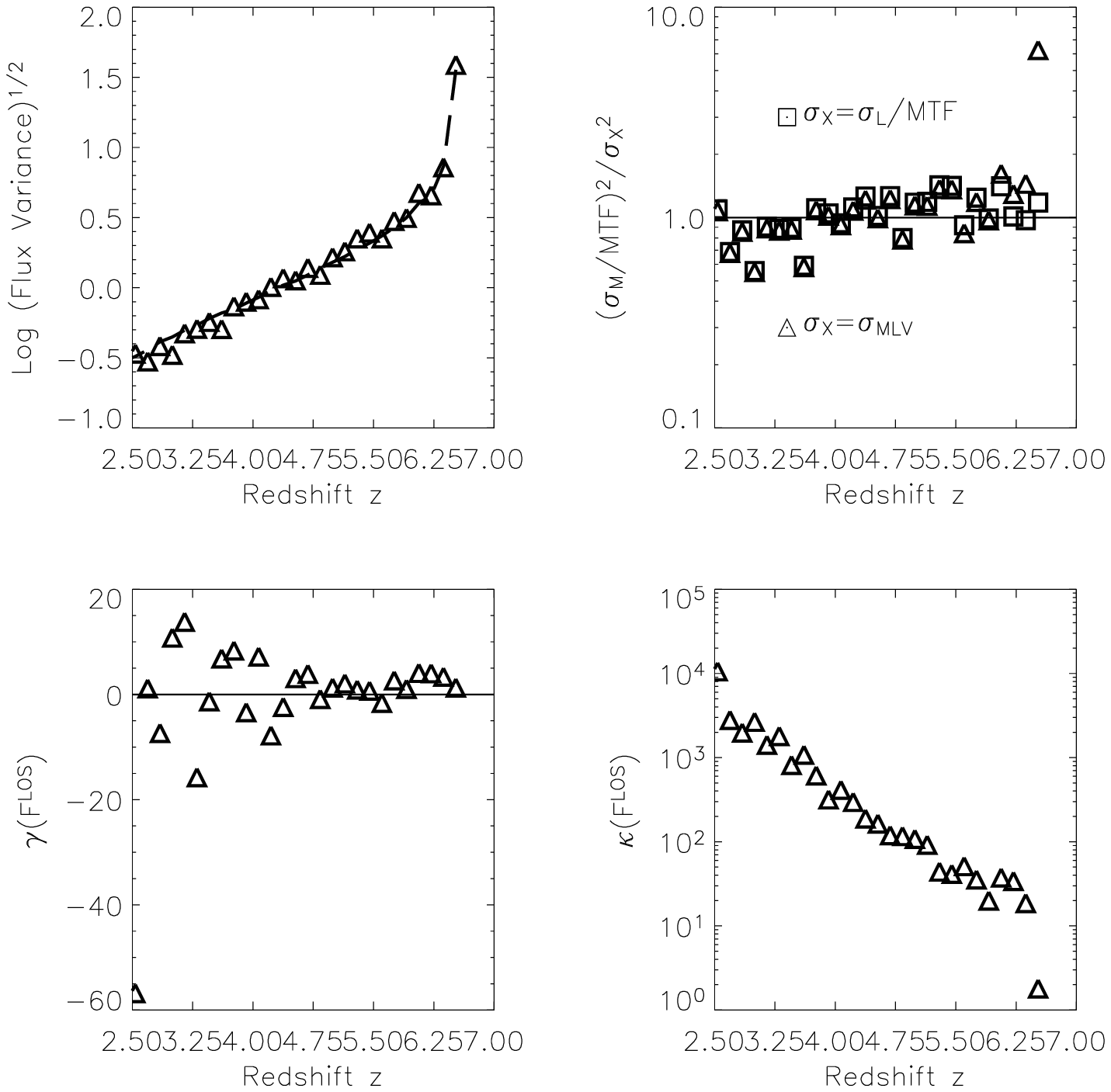,angle=0,width=6.0in,height=5.0in}
\vspace{6pt}
\figurenum{10}
\caption{
\textit{Top-left panel}    : $Log(\sigma_{M}/MTF)$ (triangles) and 
$Log(\sigma_{L}/MTF)$ (dashed line) versus redshift \newline
\textit{Top-right panel}   : $(\frac{\sigma_{M}}{MTF})^{2}/\sigma_{X}^{2}$ 
where $\sigma_{X}^{2}$ is either $\sigma_{MLV}^{2}$ (mean los-variance: triangles) 
or $(\sigma_{L}/MTF)^{2}$ (mean standard los-variance normalized to the mean transmitted
flux: squares).  \newline 
\textit{Bottom-left panel} : Skewness of the LOS-mean fluxes $F_{j}$ (j=1,NLOS)\newline
\textit{Bottom-right panel}: Kyrtosis of the LOS-mean fluxes $F_{j}$ (j=1,NLOS)
}
\label{fig10}
\addtolength{\baselineskip}{10pt}
\end{figure}

\newpage

%
\begin{figure}
\epsfig{figure=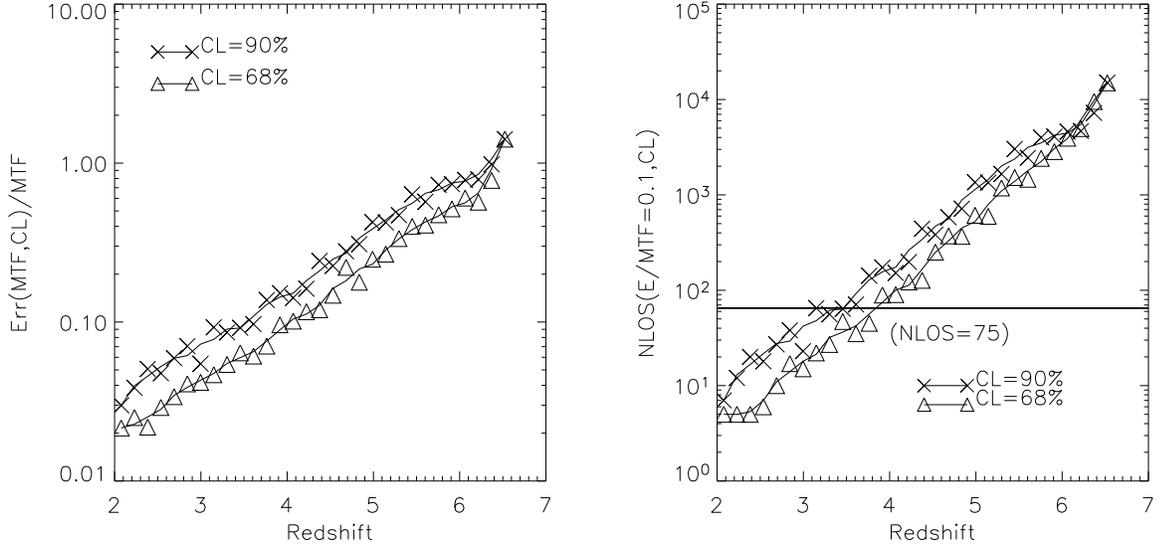,angle=0,width=6.0in}
\vspace{6pt}
\figurenum{11}
\caption{
On the left panel of Figure~(11) we plot the relative to the MTF margin of error as it scales
with redshift. At the 90\% confidence level the margin of error is $\propto 60$\%
the MTF value at $z=5.5$ ($\propto 40$\% at CL=68\%) and larger than 100\% at $z>6.4$.
This shows that 75 lines of sight undersample the distribution of mean fluxes
during reionization. If we assume that the LOS-mean flux measurements are normally distributed
at all redshifts then we can estimate the required NLOS number in order to
determine the margin of error within 10\% of the MTF. We would need
1200 lines of sight to determine the MTF within 10\% (CL=90\%) at $z~\approx~5$. The
required number of LOS drops to about 600 at the 68\% confidence level at $z~\approx~5$.
}
\label{fig11}
\addtolength{\baselineskip}{10pt}
\end{figure}

\newpage

%
\begin{figure}
\epsfig{figure=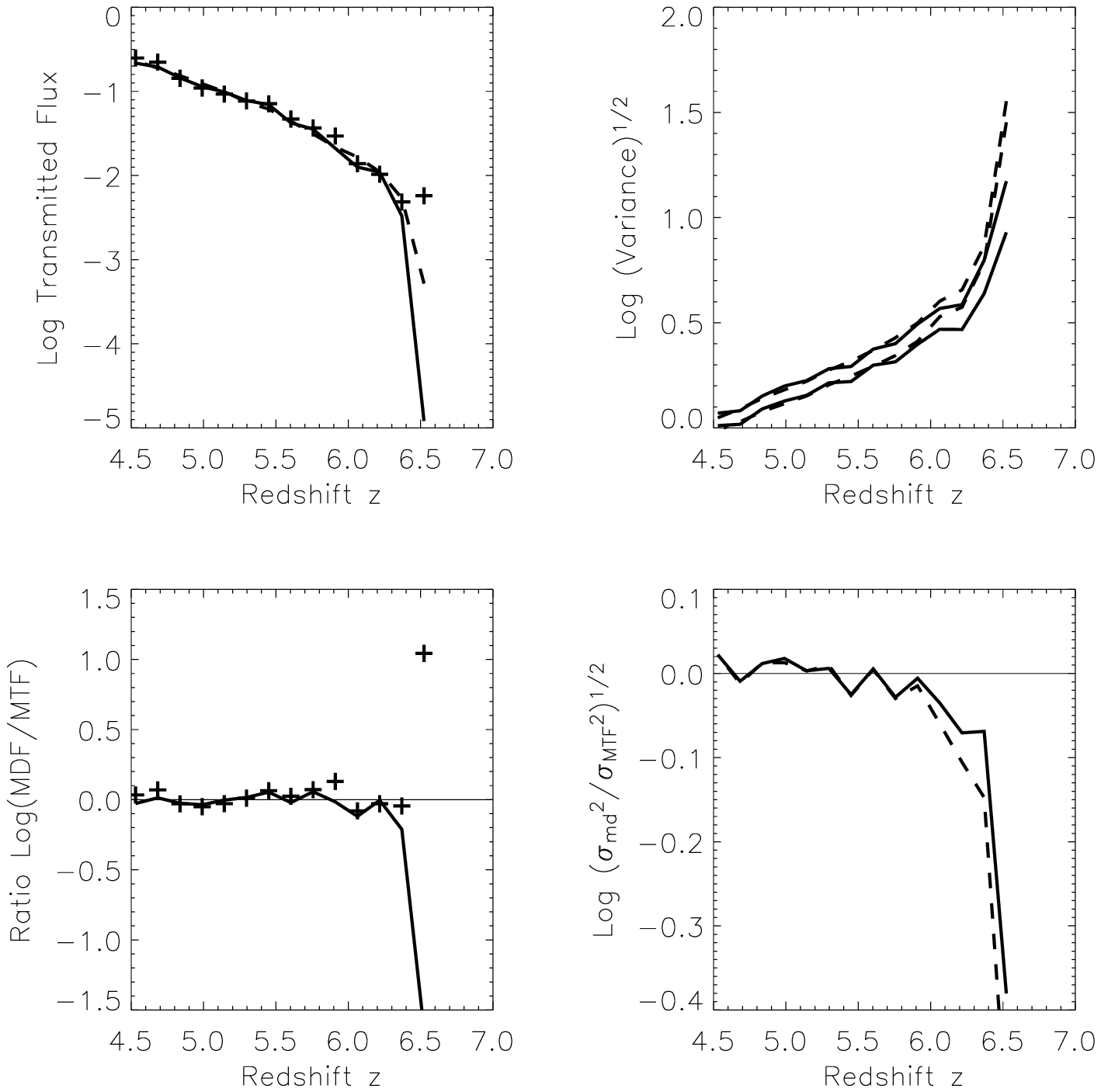,angle=0,width=6.0in,height=5.0in}
\vspace{6pt}
\figurenum{12}
\caption{
\textit{Top-left panel}: The solid line shows the redshift evolution of the MDF
compared to the redshift evolution of the MTF (dashed line). The cross points
were computed from lines of sight that have mean fluxes $>10^{-3}$ in the
last redshift interval. \newline
\textit{Top-right panel}: The solid lines are the redshift evolution of the flux
variance (upper: FRES, lower: LRES) of the lines of sight with LOS-mean values 
within the mode of the LOS-means. The dashed lines are the redshift evolution of the
MTF variance (upper: FRES, lower: LRES). \newline
\textit{Bottom-left panel}: The solid line shows the redshift evolution of the
ratio $Log_{10}(MDF/MTF)$. The cross points show the ratio between the mean flux
computed from high-z high transmission lines of sight only over the MTF at each
redshift. \newline
\textit{Bottom-right panel}: The redshift evolution of the ratio of 
$\sigma_{md}$ to $\sigma_{MTF}$ is shown in the two resolution cases (solid: FRES, dashed:
LRES).
}
\label{fig12}
\addtolength{\baselineskip}{10pt}
\end{figure}

\newpage

%
\begin{figure}
\epsfig{figure=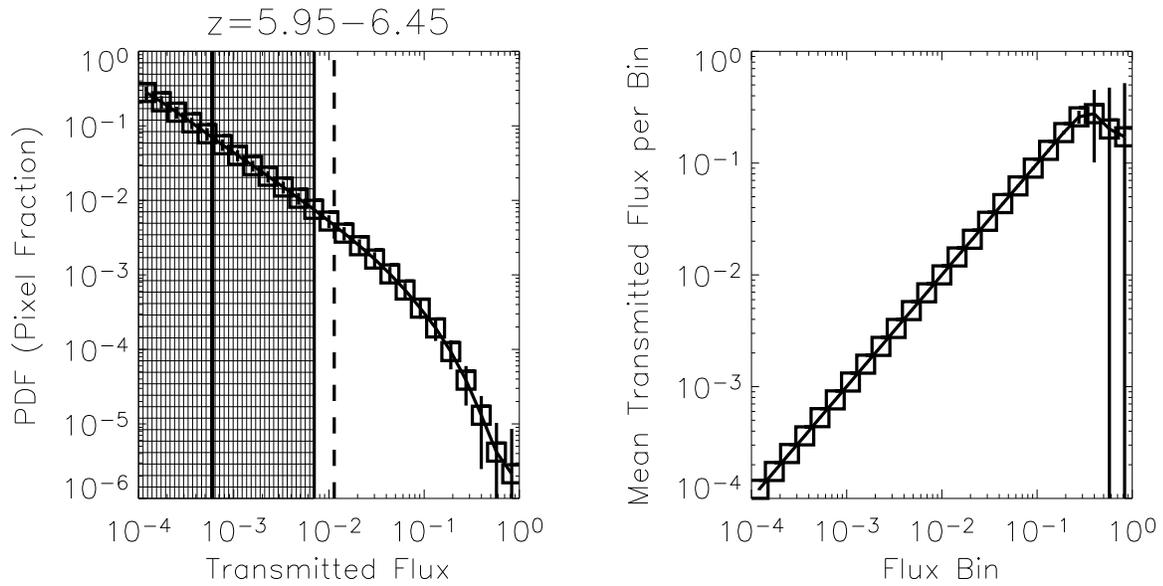,angle=0,width=6.0in,height=3.0in}
\vspace{6pt}
\figurenum{13}
\caption{
Left panel: Discrete flux distribution (DFD) in the redshift interval [5.95,6.45].
Overplotted are the mean flux of the Becker gap (solid line) and the extremes of the
transmitted flux (shaded area).
Right panel: The mean transmitted flux in each of the flux bins of the DFD curve
is plotted against the mean flux in each bin.
}
\label{fig13}
\addtolength{\baselineskip}{10pt}
\end{figure}

\newpage

%
\begin{figure}
\epsfig{figure=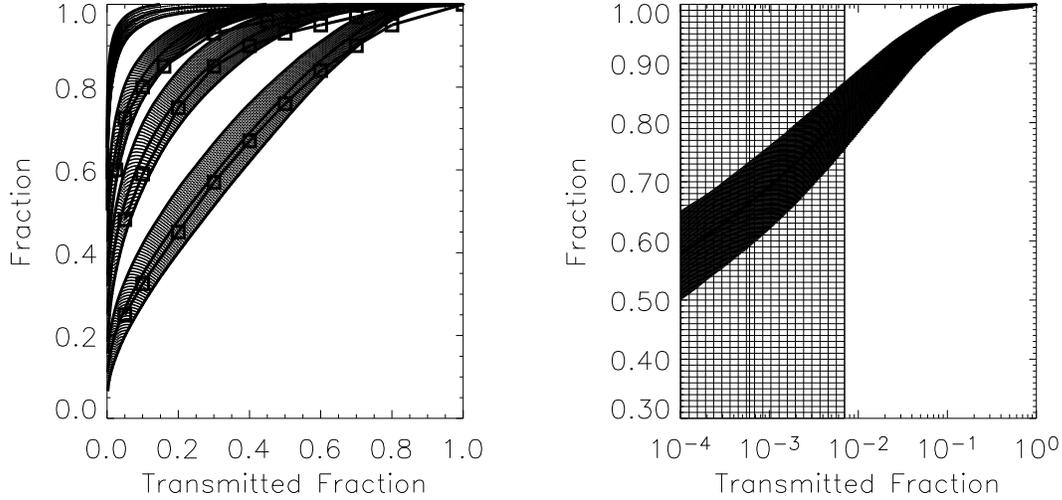,angle=0,width=6.0in,height=3.0in}
\vspace{6pt}
\figurenum{14}
\caption{
Cumulative flux distribution (CFD) in four redshift bins (from bottom up) (4.25-4.75),
(4.95-5.45),(5.45-5.95) \& (5.95-6.45). The line of sight fluxes in all z-bins were
convolved down to the LRES spectral resolution. The data points for the first three
bins are extracted from the observational distribution on Figure~(8) in Songaila et al
(2002). In each transmitted flux bin (x-axis) we average the fraction (y-axis) from
all LOS and compute the standard deviation.  The shaded areas are included between the
$\pm 2 \sigma$ curves. There is a general agreement with the observed data in the low
and medium transmission regions. Any disagreement in the high transmission end of the
x-axis has to take into consideration the sensitivity of the CFD to the extrapolated
continuum in the observed data.
}
\label{fig14} 
\addtolength{\baselineskip}{10pt} 
\end{figure}

\newpage

%
\begin{figure}
\epsfig{figure=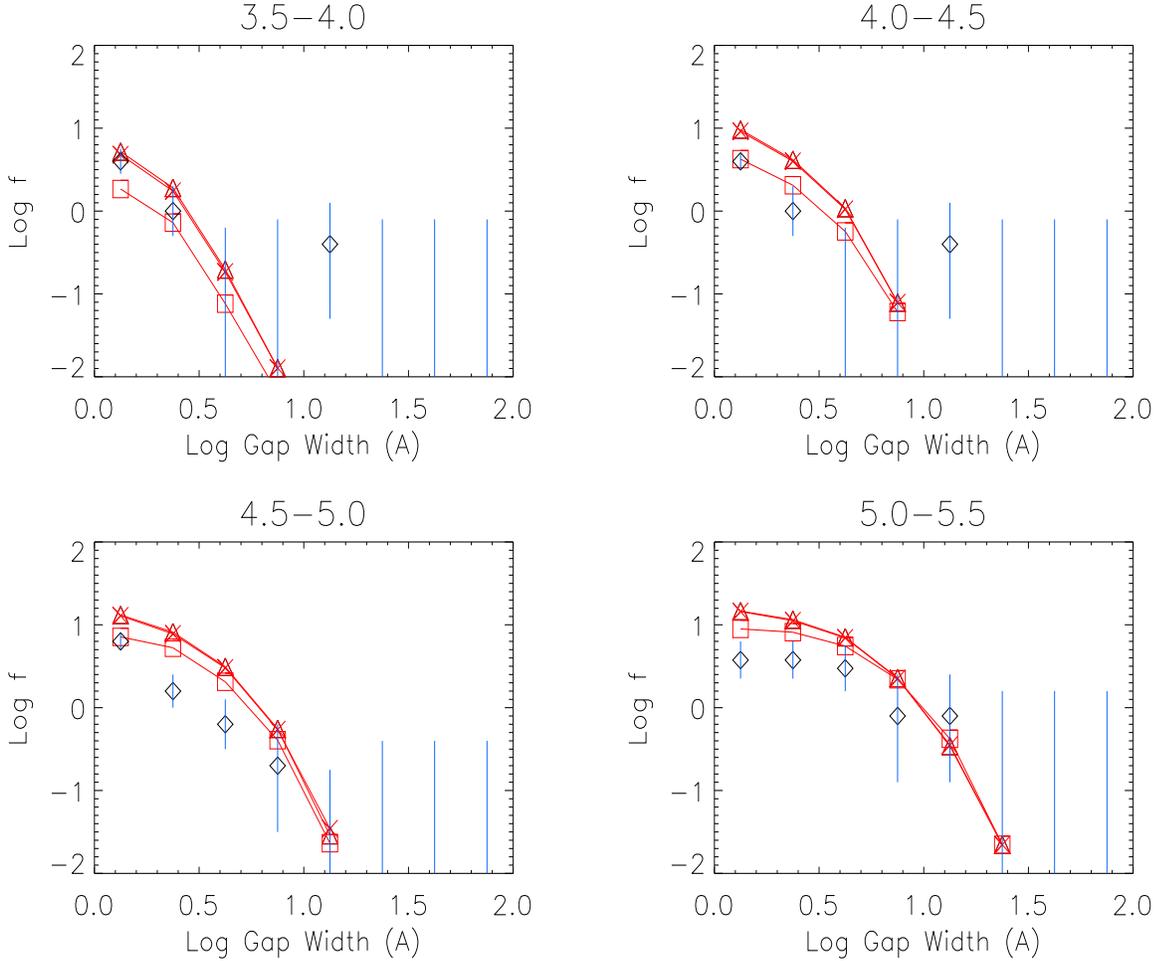,angle=0,height=5.0in,width=6.0in}
\vspace{6pt}
\figurenum{15}
\caption{
Frequency of contiguous gaps in our synthetic spectra for the
four redshift intervals examined by Songaila (2002). We plot all three spectral
resolution cases we studied (FRES: triangles, HRES: crosses, LRES: squares).
Between z=3.5-5.5 there is no evident difference between the HRES and FRES cases.
When compared to the higher resolution results the LRES case 
systematically underestimates the small gap widths 
while it systematically overestimates the large gap widths. 
In all cases, the GWD extends to larger gap-widths as the redshift
increases while it simultaneously flattens in the small gap-widths range 
(GWW $< 3$ \AA).
}
\label{fig15}
\addtolength{\baselineskip}{10pt}
\end{figure}

\newpage

%
\begin{figure}
\epsfig{figure=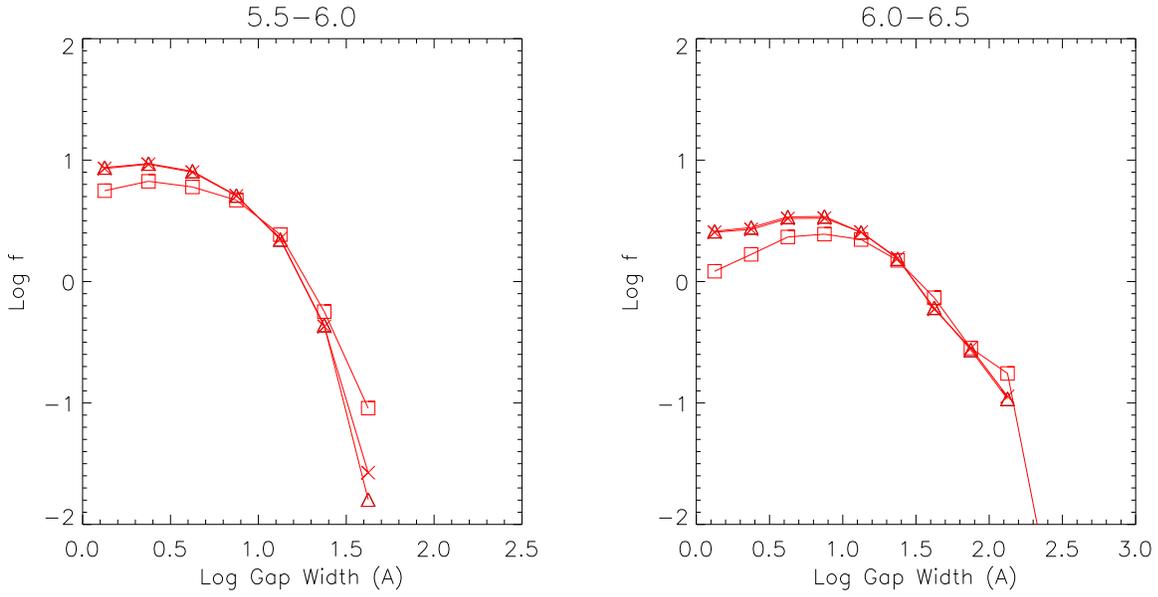,angle=0,width=6.0in}
\vspace{6pt}
\figurenum{16}
\caption{
Same as in Figure~(15) for the two high redshift intervals at $z=5.5-6.5$. 
The GWD slowly evolves between z=5-6 to larger GWW values ($>30$ \AA) and a flatter
profile at $GWW < 6$ \AA. The last point marks the progressive disappearance of
small gaps which accelerates at $z>6$. The decrease in the frequency 
of small gap sizes and the simultaneous, increase in the frequency of large gaps
suggests a mechanism of gap merging that occurs as the opacity of the IGM increases
close to reionization.
}
\label{fig16}
\addtolength{\baselineskip}{10pt}
\end{figure}

\newpage

%
\begin{figure}
\epsfig{figure=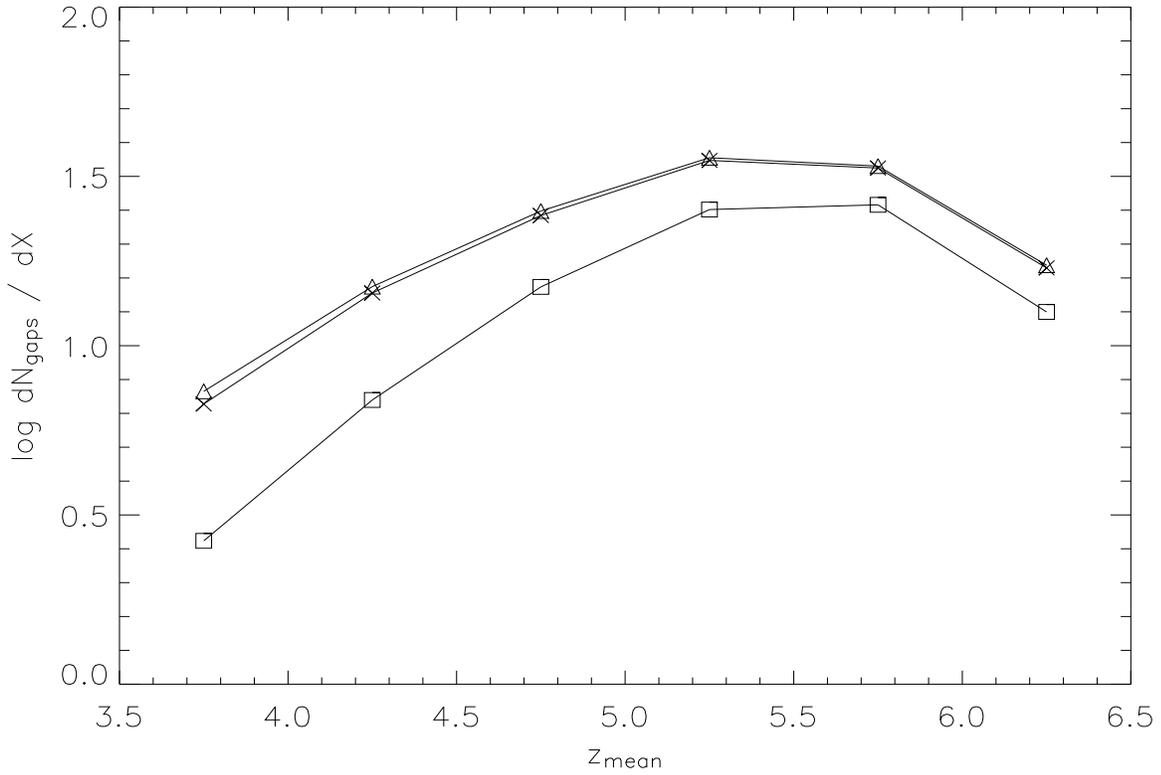,angle=0,width=6.0in,height=4.0in}
\vspace{6pt}
\figurenum{17}
\caption{
Evolution of the total number of gaps per redshift path (squares: LRES, crosses: HRES, triangles: FRES). 
Prior to entering the last stages of reionization the number of gaps peaks at $z \sim 5.5$ and then
decreases due to "gap merging". 
}
\label{fig17}
\addtolength{\baselineskip}{10pt}
\end{figure}

\newpage
\clearpage

%
\begin{figure}
\epsfig{figure=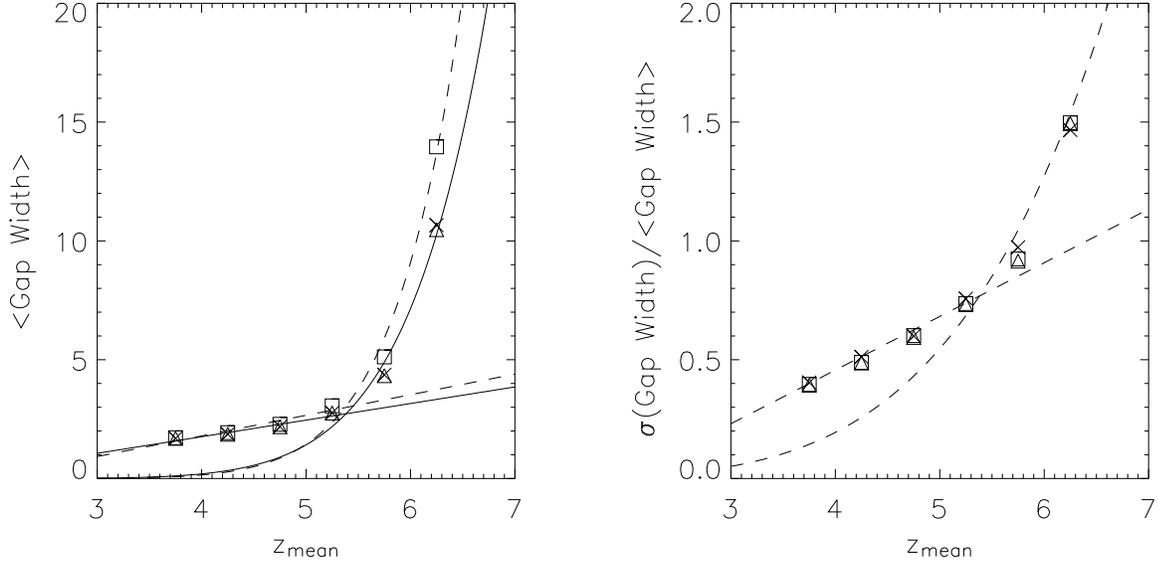,angle=0,width=6.0in}
\vspace{6pt}
\figurenum{18}
\caption{
On the left panel we plot the log of the average gap width against redshift for the
three resolutions examined. (squares: LRES, crosses: HRES, triangles: FRES). The mean
gap width increases linearly with redshift up to $z \sim 5.25$ and then rapidly evolves under
a hard power-law.
On the right panel we plot the log of the ratio 1$\sigma$ error over the 
mean gap width. Large ratios ($>1$) indicate significant scattering about the mean 
which is due to the GWD extending at larger gap widths close to the reionization phase. 
The dashed curves are fits to the data. A linear profile at $z \leq 5.25$ and a power-law
profile at $z \geq 5.75$. Cubic interpolation between the data assisted in deriving the power-law
fit.
}
\label{fig18}
\addtolength{\baselineskip}{10pt}
\end{figure}

\newpage
\clearpage

%
\begin{figure}
\epsfig{figure=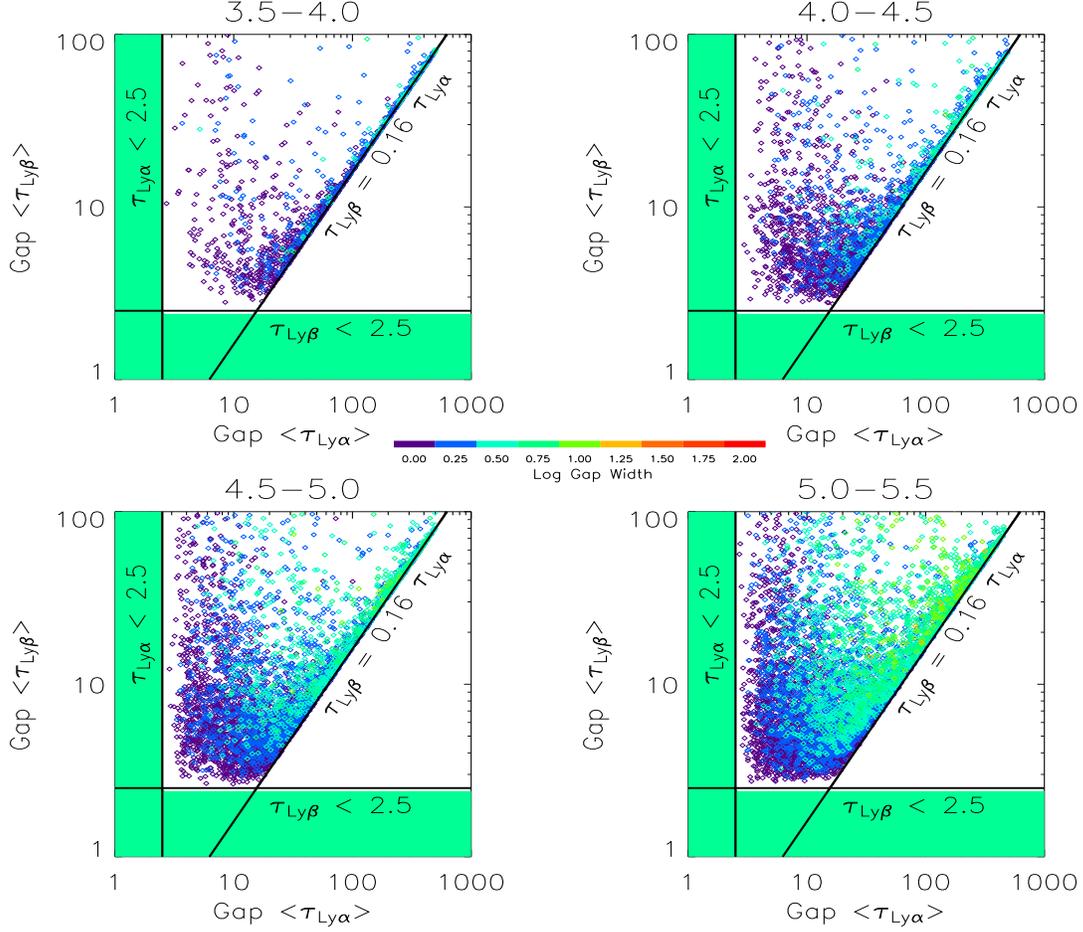,angle=0,width=6.0in,height=5.0in}
\vspace{6pt}
\figurenum{19}
\caption{
Scatter plot between the mean Ly$\alpha$ and the mean Ly$\beta$ optical depths 
for each gap measured with wavelength width spread larger than 1 \AA ~and for 
optical depth pixels greater than 2.5. The slope of the straight line is 0.16,
equal to the ratio $\frac{f_{Ly \beta } \lambda_{Ly \beta }}{f_{Ly \alpha }
\lambda_{Ly \alpha }}$.
The color table on the figure shows the allocation of color for each pair of
optical depths based on their measured gap width. The points were plotted from the 
smallest to the largest gap values and therefore the colored symbols 
represent the largest gap value measured in the locale of each optical depth pair 
$(\tau_{Ly \beta}, \tau_{Ly \alpha})$.
}
\label{fig19}
\addtolength{\baselineskip}{10pt}
\end{figure}

\newpage

%
\begin{figure}
\epsfig{figure=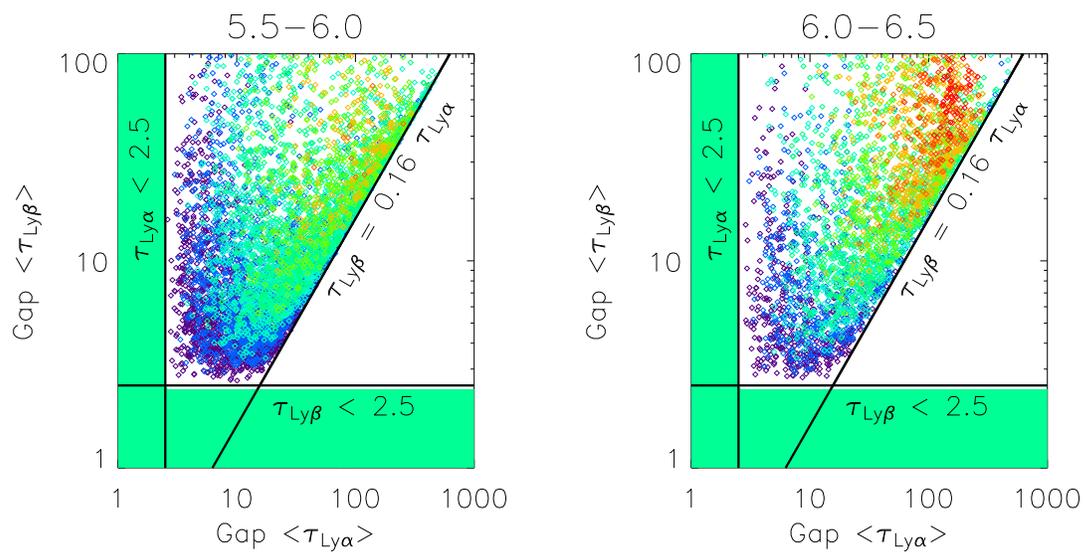,angle=0,width=6.0in,height=3.0in}
\vspace{6pt}
\figurenum{20}
\caption{
Same as in Figure~(19) but for the redshift intervals (5.5-6.0) \& (6.0-6.5). 
Gaps with wavelength widths larger than $10^{1.75} \approx 56$ \AA ~(orange color points)
and 100 \AA ~(red color symbols) become evident in each of the two redshift intervals
respectively. On the right panel we can see that the $> 100$ \AA~ gaps at $z \geq 6$ 
correspond to mean Ly$\alpha$ optical depths larger than 100.
}
\label{fig20}
\addtolength{\baselineskip}{10pt}
\end{figure}

\newpage


\begin{figure}
\epsfig{figure=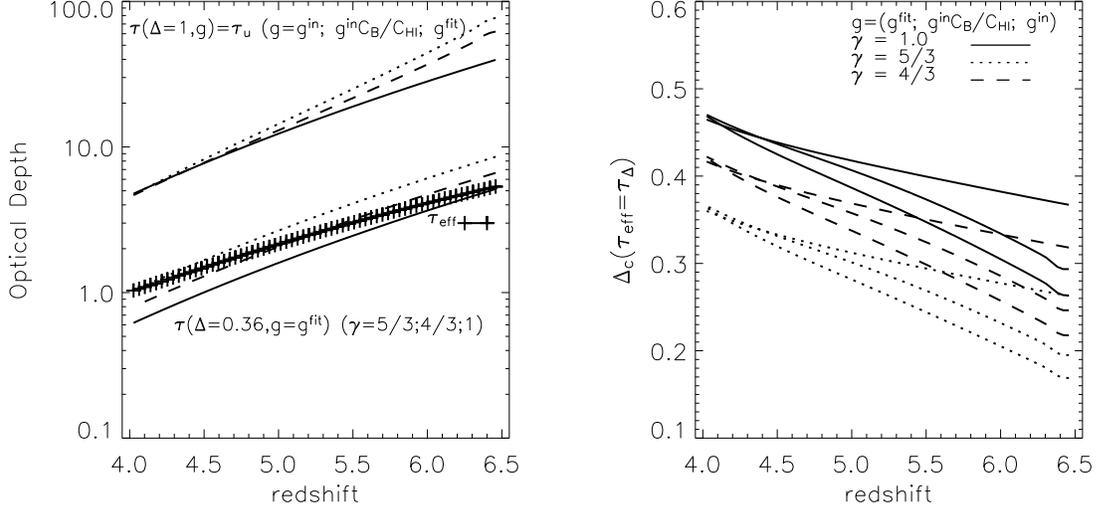,angle=0,width=6in,height=3.0in}
\vspace{6pt}
\figurenum{21}
\caption{
On the left panel we plot the redshift evolution of the optical depth 
at the cosmic mean density $\tau_{\Delta=1} = 14g^{-1}~(\frac{1+z}{7})^{4.5}$.
The three cases illustrated correspond to different choices for the normalized ionization
rate. From top to bottom: rate from raw simulation data ($g^{in}$), rate modified by the ratio 
of baryon to HII clumping factors ($g^{in} \frac{C_{B}}{C_{HII}}$)  and the rate inferred by 
the fit of our optical depth data to the functional form of Equation~(3).
Overplotted is the redshift evolution of the effective optical depth (crosses) and 
$\tau_{\Delta}=\tau_{\Delta=1}~\Delta^{\beta}$, for $\Delta=0.36$ and (from top to bottom)
$\gamma=\frac{5}{3},\frac{4}{3},1$.
On the right panel, we compute the overdensity where $\tau_{eff}=\tau_{\Delta=1}~\Delta^{\beta}$ 
in the redshift interval z=4-6.45 for the normalized ionization rate types and adiabatic indices range 
mentioned. The figure shows that at high redshifts ($z > 4$) the effective optical depth 
derivation is biased by small overdensities ($\Delta < 0.5$). 
}
\label{fig21}
\addtolength{\baselineskip}{10pt}
\end{figure}

\newpage


\begin{figure}
\epsfig{figure=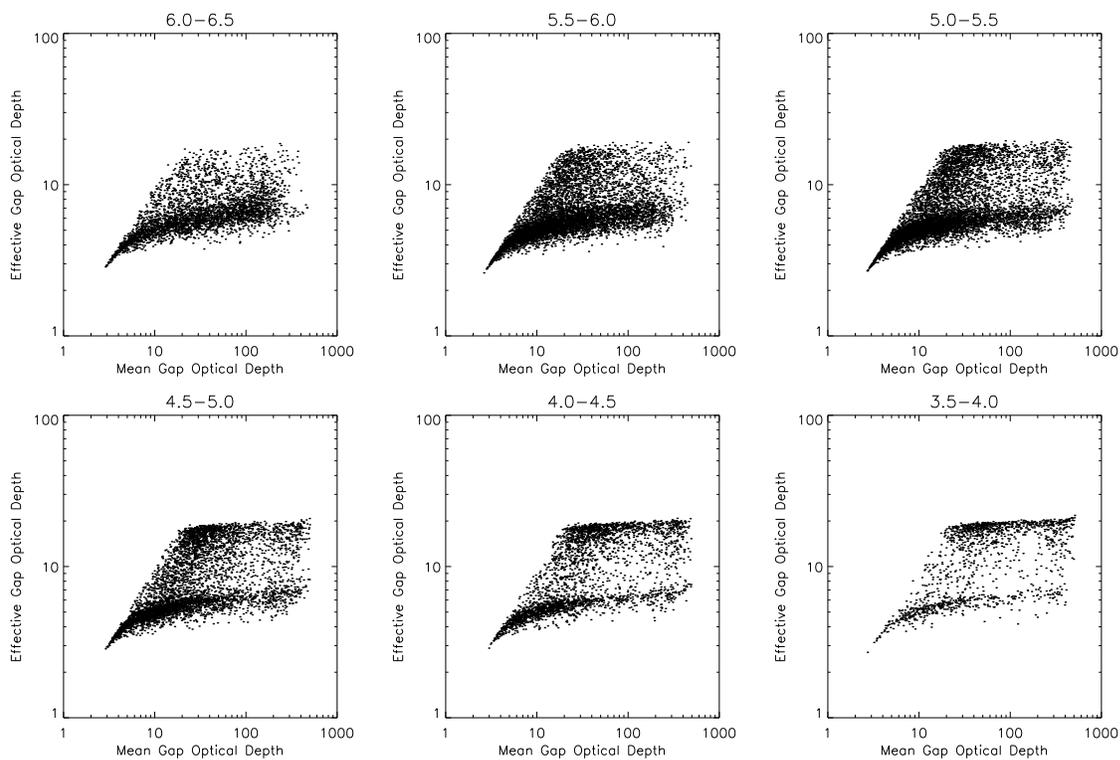,angle=0,width=6.0in,height=4.0in}
\vspace{6pt}
\figurenum{22}
\caption{
Redshift evolution of the scatter plot between the Ly$\alpha$ mean optical depth and the 
Ly$\alpha$ effective optical 
depth of dark gaps. In the range $\bar{\tau}_{gap} > 15$ the two types are not correlated at 
$z \leq 6$ and weakly correlated at $z \geq 6$. For mean optical depths 
$\bar{\tau}_{gap} \leq 15$ there is a positive correlation which becomes stronger 
as the gap optical depth decreases.
}
\label{fig22}
\addtolength{\baselineskip}{10pt}
\end{figure}

\newpage


\begin{figure}
\epsfig{figure=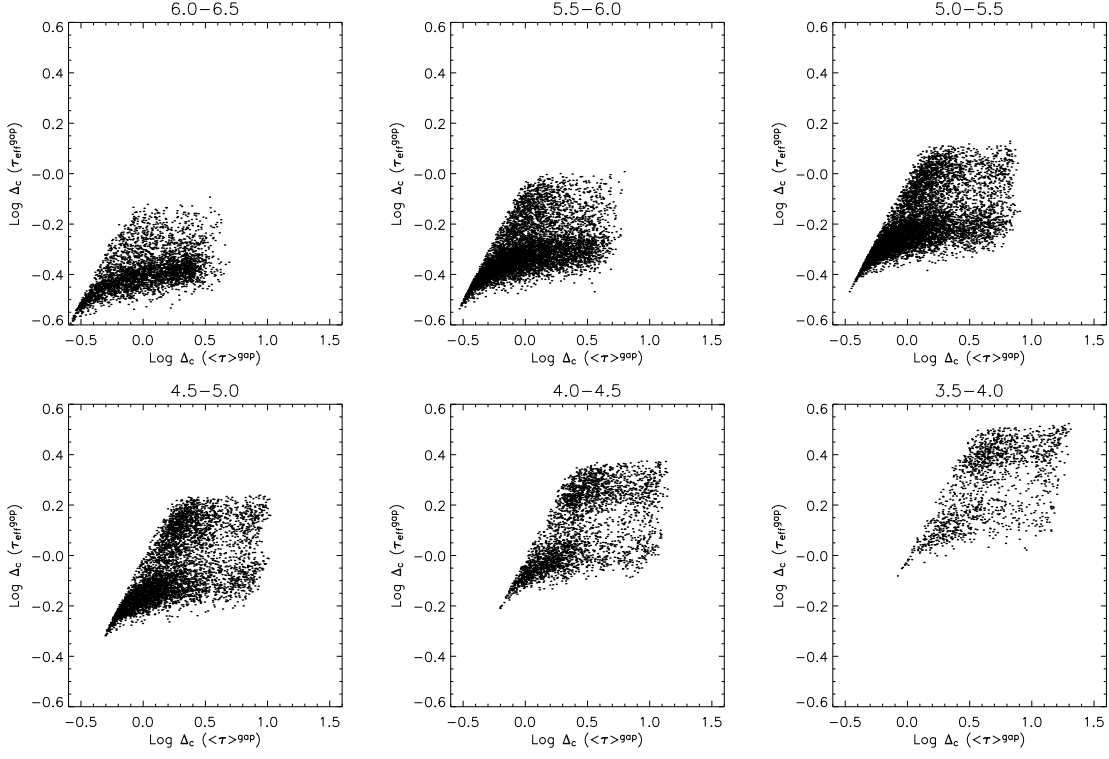,angle=0,width=6.0in,height=4.0in}
\vspace{6pt}
\figurenum{23}
\caption{
Redshift evolution of the scatter plot between the characteristic overdensity $\Delta_{c}$ and
the inferred average optical depth of a dark gap. The plots were compiled from Figure~(22)
through the equation $\Delta_{c} = (\frac{\tau}{\tau_{\Delta=1}})^{1/\beta}$ for $\gamma=4/3$ and 
$\tau=<\tau_{gap}>$ (mean optical depth) or $\tau=\tau_{eff}^{gap}$. The figure shows that the 
gap effective optical depth is biased by smaller overdensities than the gap mean optical depth.
As we approach reionization ($z \geq 6$) $\Delta_{c}^{eff}$ is entirely in the 
underdense ($\Delta < 1$) range. 
}
\label{fig23}
\addtolength{\baselineskip}{10pt}
\end{figure}

\newpage

\begin{figure}
\epsfig{figure=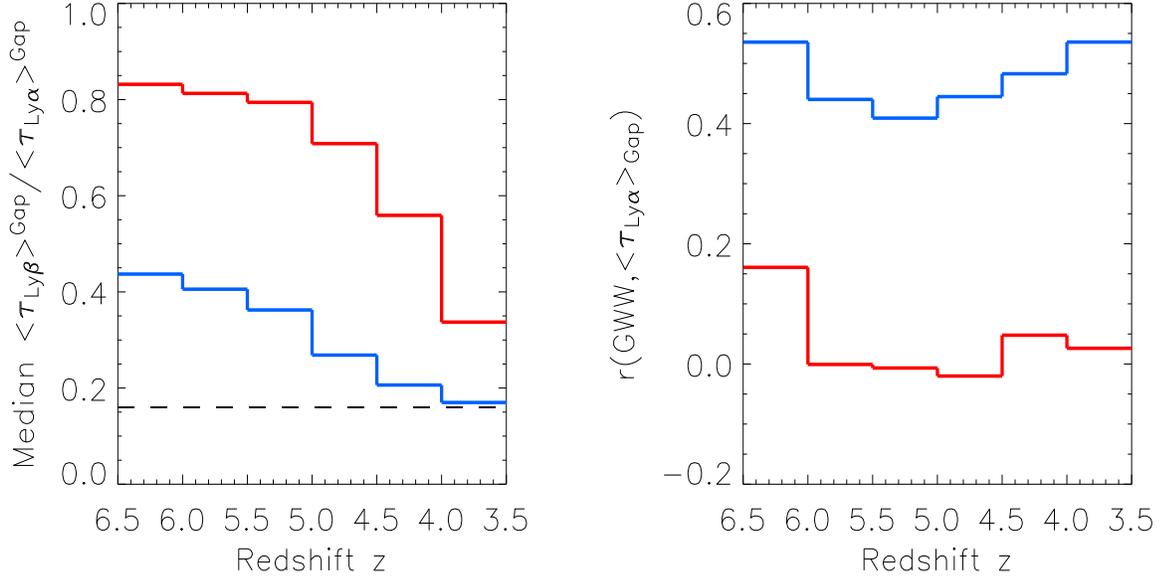,angle=0,width=6.0in,height=3.0in}
\vspace{6pt}
\figurenum{24}
\caption{
Left Panel: Evolution of the median 
ratio between the Ly$\beta$ and Ly$\alpha$ optical depths from Figures~(19,20) (blue 
histogram). For comparison, we overplot (red histogram) the median ratio 
derived using the gap effective optical depth. The ratio reflects an unbiased statistical
relationship of the Ly$\beta$ to Ly$\alpha$ optical depth scatter inferred from Equation~(5). 
As the redshift decreases both ratios decrease toward the expected asymptotic value (0.16) 
which in the case of the mean optical depth is reached by $z \sim 3.5$. \newline
Right Panel: Evolution of the Pearson Correlation Coefficient between the distributions of
the gap Ly$\alpha$ optical depth and the gap wavelength width (GWW).
A "weak" ($r \sim 0.5$) but positive correlation is inferred between the gap mean optical depth
and the gap size across the redshift range $3.5 \leq z \leq 6.5$. In contrast, the
the gap effective optical depth and gap size are uncorrelated at $z \leq 6$. A much weaker
correlation, when compared to the mean optical case, is seen in the redshift interval
$6 \leq z \leq 6.5$.
}
\label{fig24}
\addtolength{\baselineskip}{10pt}
\end{figure}

\end{document}